\documentclass[12pt,english]{article}
\usepackage{geometry}
\usepackage{graphicx}
\usepackage{amsmath}
\usepackage{hyperref}
\usepackage{amssymb}
\usepackage{mathtools}
\usepackage{xfrac}
\usepackage{breqn}
\usepackage{amsfonts}
\usepackage{graphicx,epstopdf}
\usepackage{subfigure}
\usepackage{fullpage}
\usepackage{float}

\hypersetup{
	colorlinks=true,
	linkcolor=blue,
	filecolor=magenta,
	urlcolor=cyan,
	citecolor=green,
	pdfpagemode=Fullscreen}
\begin{document}
\begin{center}{\Large \textbf {Study of Breather Structures in the Framework of Gardner Equation in Electron-Positron-Ion Plasmas}}
\end{center}
{\large \par} \vspace*{.5in}
\begin{center}\textbf{Snehalata Nasipuri$^{1}$,  Swarniv Chandra$^{2}$, Uday Narayan Ghosh$^{3*}$, Chinmay Das$^{4}$, Prasanta Chatterjee$^{1}$ }\\
\textbf{1. Department of Mathematics,} \textbf{Visva-Bharati, Santiniketan-731235, India}\\
\textbf{2. Department of Physics, Government General Degree College at Kushmandi, Dakshin Dinajpur-700032, India}
\\
\textbf{3. Department of Mathematics, K. K. M. College,} \\\textbf{A Constituent unit of Munger University, Jamui, Bihar-811307, India}\\
\textbf{4. Department of Mathematics, K.J.R. Government General Degree College, Bankura-722143, India}
\end{center}
\vspace*{.25in}
\begin{center}
Abstract
\end{center}
In different nonlinear mediums, the wave trains carry energy and expose many amazing features. To describe a nonlinear phenomenon, a soliton is one that preserves its shape and amplitude even after the collision. Breather is one kind of soliton structure, which is a localized wave that periodically oscillates in amplitude. In the Korteweg-de Vries (KdV) equation, nonlinearity appears in quadratic terms, whereas the modified Korteweg-de Vries (mKdV) equation contains cubic nonlinearity. When we consider both the quadratic and cubic nonlinearities, we obtain the Gardner equation (GE). So GE is also called the KdV-mKdV equation. Due to the prominent balance between nonlinear terms (quadratic and cubic) and higher-order linear terms in an evolution equation, one can get a soliton profile solution, but when both (quadratic and cubic) nonlinear terms and higher-order linear terms appear together in an evolution equation, we are enabled to obtain breather soliton structures.  This article uses the reductive perturbation technique (RPT) to get the GE from a plasma system with four parts: cold positrons that can move, hot positrons and hot electrons that are spread out in a kappa pattern, and positive ions that can't move. Then, using the Hirota bilinear method (HBM), it is possible to obtain the multi-soliton and breather structures of GE. Breathers are fluctuating regional wave packets and significantly participate in hydrodynamics as well as optics; besides, their interaction can alter the dynamical characteristics of the wave fields. We also incorporate a detailed numerical simulation study based on a newly designed code by two of the co-authors. It is found that in our plasma system, soliton solutions, especially breather solutions, exist. Although superthermal (kappa-distributed) electrons and positrons play an important role in soliton structures, This type of analysis can also apply to the propagation of finite-amplitude waves in natural phenomena like the atmosphere, ocean, optic fibres, signal processing, etc. It should also be useful to study different electrostatic disturbances in space and laboratory plasmas, where immobile positive ions, superthermal electrons, superthermal hot positrons, and mobile cold positrons are the major plasma species.\\\\
\textbf{Keywords}:  Positron acoustic waves; Kappa distribution; Hirota bilinear method; modified Korteweg-de Vries; Gardner equation; Solitons; Breathers.
\vspace*{1in}

*Corresponding author e-mail: unghosh@gmail.com\\
\section{Introduction}
Nowadays, the study of nonlinear phenomena in electron-positron (e-p) and electron-positron-ion (e-p-i) plasmas attracts more attention from researchers due to their existence in astrophysical space plasmas as well as laboratory plasmas like supernovas, quasar and pulsar magnetospheres, cluster explosions, active galactic nuclei, Van Allen radiation belts, polar cups of fast rotating neutron stars, semi-conductor plasmas, and intense laser fields. In linear theory, the amplitude of the wave is sufficiently small to neglect the contributions of second- and higher-order terms, i.e., nonlinear terms. In the case of larger wave amplitudes, the linear approximation breaks down and nonlinearity is taken into account. The harmonic generation, including fluid advection, the nonlinear Lorentz force, the trapping of particles in the wave potential, etc., introduces the nonlinearities. In plasmas, the nonlinear effect leads to localised waves and creates different types of structures like soliton solutions, breather solutions, etc. Our main object is to find the soliton structures and the breather structures for a four-component plasma system consisting of immobile positive ions. mobile cold positrons and Kappa-distributed hot positrons and hot electrons, and we are able to establish such observations by deriving the Gardner equation (GE) from the above-mentioned model equations. GE is used to model the wave propagation in a one-dimensional nonlinear lattice with a nonharmonic-bound particle. Since GE is an extension of KdV and mKdV equations, the properties of GE are obtained by the properties of the mKdV equation. In this article, we have studied a special nonlinear feature, which is the breather soliton structure. We know that solitary waves are localised, travelling steady-profile solutions for dispersive nonlinear dynamical systems. Breathers are also localised travelling waves, but their profile is not steady and changes periodically in time; they are a mix of soliton and periodic solutions. In a coordinate system moving with the wave, the wave profile is itself a moving periodic travelling wave, containing an amplitude envelope that keeps it localised in space. It is hard to characterise breathers analytically, even in (1+1)-dimension, since they are solutions for which the governing equations can't be reduced to a lower-order system. So, to produce breathers in weakly nonlinear limits, where breathers are of small amplitude, one can look at them with very shallow envelopes; in this case, the separation of scales allows a reduction of the governing equations to a lower-order system. So, our study will be helpful for further investigation of the nonlinear structures (soliton, breathers) for other evolution equations that can be derived from the mentioned system. 
\\ The high-energy particles can be found in space plasmas or laboratory plasmas, and such high energization may arise due to the effect of external forces acting on the natural space environment plasmas or wave-particle interaction. Plamas with superthermal electrons, or positrons, are generally characterised by a long tail in the high-energy region. A generalised Lorentzian or kappa distribution is used to model such space plasmas \cite{sir-book}. Many researchers have studied the nonlinear propagation of waves in e-p-i plasmas \cite{epi-1}-\cite{epi-3}. A study by Alam et al. \cite{alam-etal} looked at the effects of superheated electrons and positrons on positron-acoustic single waves and double layers in e-p-i plasmas. A three-dimensional generalised Lorentzian or kappa distribution function \cite{lorenz-kappa} can be written as:
\[ F_{\kappa}(\nu)=\frac{\Gamma(\kappa +1)}{(\pi \kappa {\theta}^2)^{(3/2)} \Gamma(\kappa -1/2) }(1+\frac{{\nu}^2}{\kappa \theta^2})^{-(\kappa +1)},\] 
where $\Gamma$ is the gamma function, $\theta$ is the most probable speed or effective thermal speed connected to the usual thermal velocity $V_t=(\frac{k_B T}{m})^{(1/2)}$ by $\theta=\frac{2 \kappa -3}{\kappa} V_t,$ $T$ is the characteristic kinetic temperature, and $k_B$ is the Boltzmann constant. The parameter $ \kappa $ is the spectral index parameter that defines the strength of the superthermality. The range of this parameter is $3/2 < \kappa< \infty$. In the limit, $\kappa \to \infty,$ the kappa distribution function reduces the well-known Maxwell-Boltzmann distribution. So, the kappa distribution function is a more generalized function than the Maxwellian distribution function. Kappa distribution has been used to interpret spacecraft data in the Jupiter and Saturn \cite{jupiter}, Earth's magnetospheric plasma sheet \cite{earth-sheet}, and the solar wind, to explain the velocity filtration effect in the solar corona \cite{scorona1}, and to analyze the field-aligned conductance values in the auroral region by using the Freja satellite data \cite{scorona2}. \\ 
From the initial works on the early universe to present-day laboratory experiments, electron-positron ion (e-p-i) plasma has been a fundamental field of interest for theoretical and experimental plasma physicists. E-P-I plasma has successfully explained where gamma-ray sources come from and how they work by combining relativistic electrons and positrons in a strong laser field \cite{ref1}. Electrons and positrons form an inverted, two-level system. In such types of studies, Klein-Gordan field theory can be helpful. Later in 2006, Eliasson and Shukla \cite{eliasson} studied coherent structures in phase-space vortices in pair plasma. Pitrou et al. \cite{pitrou} have given a picture of the Big Bang nucleosynthesis, including stages of e-p-i plasma. \\
Luque and Schamel \cite{luque} have theorised that electrostatic trapping can be a key to the dynamics of plasmas and other similar collective systems. Volker and Tsupko \cite{volker} have carried out a detailed calculation and showed how the e-p-i plasma influences the black hole shadows. These works also refer to previous works of Breuer and Ehlers \cite{breuer}. Ruffini et al. \cite{ruffini} have referred to electron-positron (e-p) pairs in physics, from heavy nuclei to black holes. This work refers to gravitational collapse leading to the formation of black holes observed in gamma-ray bursts (GRBs). Even studies in new fields like the generation of antihydrogen have electron-positron plasma \cite{antihydro}. Such a study was instrumental in many stages of research on the antiproton decelerator at CERN in Geneva. Thus, we see that the study of antimatter also involved references to the correlative properties of e-p-i plasma (\cite{cor-epi1},\cite{cor-epi2}). This work reviews the recombination mechanisms of electrons and positrons to produce antihydrogen. The ATHENA and ATRAP collaborations with CERN have dealt a lot with such e-p pair plasma. \\
Coming to works of interest we have found how ion acoustic shock waves in plasmas are formed with warm ions and Kappa-distributed electrons and positrons (Hussain et al. \cite{hussain}). Mehdian \cite{mehdian} and collaborators have studied ion-acoustic waves carrying orbital angular momentum (OAM) in the e-p-i plasmas. Asit Saha \cite{asit} studied the nonlinear excitations for the positron acoustic shock waves in dissipative nonextensive unmagnetized e-p-i plasmas. With a magnetized e-p-i plasma two-stream instability has been studied by Tinakiche et al. \cite{tinaki}. Moslem et al. \cite{moslem} and others \cite{sarkar} studied nonlinear excitations in e-p-i plasmas in accretion disks of active galactic nuclei (AGN). Now, non-Maxwellian electron distribution plays an important role. It may be Cairsis distributed or super thermally distributed (Kappa, Shamel, etc). Jilani \cite{jilani}  has shown how electron acoustic waves (EAWs) evolve in magneto-rotating e-p-i plasma with nonthermal electrons and positrons. \\ 
The effect of dust in e-p-i plasmas with superthermal electrons has been also studied by Siberian et al. \cite{siberian}. Mugemana et al. \cite{mugemana} have incorporated charge separation in nonlinear waves in e-p-i plasmas. Furthermore, gyrokinetic stability analysis in e-p-i plasmas was carried out by Mishchenko et al. \cite{mishchenko}. The effect of trapped ions was studied by Alinejad and his team \cite{alinejad}. Two-dimensional electrostatic shock waves in relativistic e-p-i plasma were studied by Masood \cite{masood}. Heavy ion-acoustic rogue waves in plasmas, which contain electrons, positrons, and positive and negative ions, were studied by Chowdhury et al. \cite{chowdhury}. \\
Maroof \cite{maroof} have studied magneto-hydrodynamic (MHD) waves in e-p plasmas with electron spin. Even non-planar cylindrical fast magneto-sonic solitary waves in degenerate e-p-i plasma were studied by Abdikian \cite{abdikian}. 
Recently, Kaur et al. \cite{kaur} have studied the ion-acoustic breathers in electron-beam plasma. The breather structures and peregrine solitons in a polarized space dusty plasma have been investigated by Saini et al. \cite{saini}. Slathia et al. have shown that dust-acoustic solitons, rogons, and breathers exist in Jupiter's magnetospheric dusty plasmas \cite{slathia}.\\

The motivation behind carrying out this work is the study of field quantities in an electron-positron-ion plasma over the gradual evolution of stationary structures. In the past numerous works have been done in lines of regular KdV equations with various plasma compositions under varied configurations. None of those works were complete in terms of the field evolutionary mechanism and its dependence on parameters. As per our knowledge, no work on breather using HBM considering our proposed model equations has been done in plasma till today. Since any intermediary stages are not possible to study analytically, probably the early authors have missed this point. So we have attempted to study the evolution by numerical means. 
\par The paper is organised in the following way: Section \ref{math-model}, contains the considered model equations. In section \ref{evo}, we present the evolutionary equations starting from the derivation of KdV, modified KdV, and finally the Gardner equation and provide their analytic solutions in the next section \ref{analytic}. The basic single soliton solution is presented in subsection \ref{1sol}, double soliton solutions in subsection \ref{2sol}, and the breather solution in subsection \ref{breathersol}. In section \ref{results}, we present the analytical (\ref{results_analy}) and numerical (\ref{results_num}) results and finally put the summary in the conclusion \ref{concl}.
	
\section{The mathematical model }\label{math-model}
We consider the non-linear propagation of positron-acoustic waves in a four-component plasma system consisting of immobile positive ions, mobile cold positrons and Kappa distributed hot positrons and hot electrons. Hence at equilibrium $n_{e0}=n_{pc0}+n_{ph0}+n_{i0}$, where $n_{i0}$, $n_{e0}$ are the unperturbed ion number density and electron number density respectively. $n_{pc0}(n_{ph0})$ is the number density of unperturbed cold(hot) positron. The hot electrons and the hot positrons are assumed to obey Kappa distribution on the positron-acoustic wave time scale and are given by the following expressions: \[ n_e=n_{e0}\left[ 1- \frac{e \phi}{K_B T_e (\kappa_e -\frac{3}{2})}\right]^{(-\kappa_e +\frac{1}{2})}, \] \[ n_{ph}=n_{ph0} \left[ 1+ \frac{e \phi}{K_B T_{ph}(\kappa_p -\frac{3}{2})} \right]^{(-\kappa_p +\frac{1}{2})}, \] where $n_e$, $n_{ph}$ are the number densities, while $T_e$ and $T_{ph}$ are the temperatures of electrons and hot positrons respectively. The real parameters $\kappa_e$ and $\kappa_p$ are the super-thermal parameters of hot electrons and hot positrons respectively. \\
The normalized basic equations governing the dynamics of positron-acoustic waves are given in dimensionless variables as follows :  
\begin{equation}\label{n1}
\frac{\partial n_{pc}}{\partial t}+ \frac{\partial (n_{pc}u_{pc})}{\partial x}=0,
\end{equation}
\begin{equation}\label{n2}
\frac{\partial u_{pc}}{\partial t}+ u_{pc}\frac{\partial u_{pc}}{\partial x}= - \frac{\partial \phi}{\partial x},
\end{equation}
\begin{multline}\label{n3}
\frac{\partial^2 \phi}{\partial x^2}= -n_{pc}-\mu_{ph}\left(1+\frac{\sigma_1 \phi}{\kappa_p -\frac{3}{2}}\right)^{(-\kappa_p + \frac{1}{2})}  + \mu_e \left( 1- \frac{\sigma_2 \phi}{\kappa_e -\frac{3}{2}} \right)^{(-\kappa_e+ \frac{1}{2})} - \mu_i ,
\end{multline}
where $n_{pc}$ is the cold positron number density normalized by its equilibrium value $n_{pc0}$, $u_{pc}$ is the cold positron fluid speed normalized by $C_{pc}= [\frac{K_B T_{ef}}{m_p}]^{(1/2)}$, $\phi$ is the electrostatic wave potential normalized by $\frac{K_B T_{ef}}{e}$, 
\[ \sigma_1 =\frac{T_{ef}}{T_{ph}}, \sigma_2= \frac{T_{ef}}{T_e}, \mu_{ph}=\frac{n_{ph0}}{n_{pc0}}, \mu_e=\frac{n_{e0}}{n_{pc0}}, \mu_i=\frac{n_{i0}}{n_{pc0}}, \]
$ T_{ef}=\frac{T_e T_{ph}}{(\mu_e T_{ph}+ \mu_{ph}T_e)} $
is the effective temperature, $K_B$ is the Boltzmann constant. $m_p$ is the positron mass, $e$ is the magnitude of the electron charge.  \\
Eq. \eqref{n2} represents the equation of motion of cold positron. As we assume the Debye length is $\gg$ gyrofrequency, the effect of all non-electrostatic fields including the magnetic field is very negligible. 
The time variable $t$ is normalized by $\omega^{-1}_{pc}= \left[ \frac{m_p}{4 \pi n_{pc0} e^2} \right]^{\frac{1}{2}}$ and the space variable $x$ is normalized by the Debye length $\lambda_{Dm}= \left[ \frac{K_B T_{ef}}{4 \pi n_{pc0}e^2} \right]^{\frac{1}{2}}$.  At equilibrium
\[\mu_{ph}=\frac{n_{ph0}}{n_{pc0}}=\frac{n_{e0}}{n_{pc0}}- \frac{n_{i0}}{n_{pc0}}-1= \mu_e -\mu_i -1. \]  
Simplifying \eqref{n3} we get 
\begin{dmath}\label{n4}
\frac{\partial^2 \phi}{\partial x^2}= -n_{pc}-\mu_{ph} \left[ 1- a_1 \sigma_1 \phi+ \frac{a_2}{2}\sigma^2_1 \phi^2 -\frac{a_3}{6}\sigma^3_1 \phi^3+ \dots \right]+ \mu_e \left[ 1+ b_1 \sigma_2 \phi + \frac{b_2}{2}\sigma^2_2 \phi^2 + \frac{b_3}{6} \sigma^3_2 \phi^3 + \dots  \right] -\mu_i ,
\end{dmath} 
where \[a_1= \frac{\kappa_p - \frac{1}{2}}{\kappa_p - \frac{3}{2}},\; b_1=\frac{\kappa_e-\frac{1}{2}}{\kappa_e -\frac{3}{2}}, \] \[  a_2= \frac{(\kappa_p -\frac{1}{2})(\kappa_p + \frac{1}{2})}{(\kappa_p -\frac{3}{2})^2},\; b_2=\frac{(\kappa_e-\frac{1}{2})(\kappa_e+ \frac{1}{2})}{(\kappa_e-\frac{3}{2})^2}, \] \[ a_3=\frac{(\kappa_p-\frac{1}{2})(\kappa_p + \frac{1}{2})(\kappa_p+ \frac{3}{2})}{(\kappa_p-\frac{3}{2})^3}, \; b_3=\frac{(\kappa_e-\frac{1}{2})(\kappa_e+\frac{1}{2})(\kappa_e+\frac{3}{2})}{(\kappa_e-\frac{3}{2})^3}. \]  \\\\
To linearise Eqs. \eqref{n1}-\eqref{n3}, we consider the dependent variable as the sum of equilibrium and perturbed parts, so, we can write $n_{pc}=1+\Bar{n_{pc}}, u_{pc}=\Bar{u_{pc}}, \phi=\Bar{\phi}.$ Here the values of parameters at the equilibrium position are given by $n_{pc}^{(1)}=1, u_{pc}^{(1)}=0, \phi^{(1)}=0.$  Hence we obtain the linearized forms of Eqs. \eqref{n1}-\eqref{n4} are given respectively as follows:
\begin{equation}\label{n11}  \frac{\partial\Bar{n_{pc}}}{\partial t}+ \frac{\partial\Bar{u_{pc}}}{\partial x}=0,
\end{equation}
\begin{equation}\label{n21}
\frac{\partial\Bar{u_{pc}}}{\partial t}+ \frac{\partial\Bar{\phi}}{\partial x}=0,
\end{equation}
\begin{equation}\label{n31}  \frac{\partial^2\Bar{\phi}}{\partial x^2}=- \Bar{n_{pc}}+\mu_{ph}a_1 \sigma_1 \Bar{\phi} +\mu_{e}b_1 \sigma_2 \Bar{\phi},
\end{equation} 
To obtain dispersion relation for low-frequency wave, we now assume that the perturbation is proportional to $e^{i(k x-\omega t)}$ and of the form \[ \Bar{n_{pc}}=n_0 e^{i(k x-\omega t)},  \Bar{u_{pc}}=u_0 e^{i(k x-\omega t)},  \Bar{\phi}=\phi_0 e^{i(k x-\omega t)}. \]
Substituting these values in the above-linearized equations, we get, \[ -i n_0 \omega +i k u_0=0,\]
\[ -i u_0 \omega +i k \phi_0=0,\]
\[ n_0 - (k^2 + u_{ph} a_1 \sigma_1 +u_{e} b_1 \sigma_2)=0.\]
Since the above three equations are a system of linear homogeneous equations, so for nontrivial solutions we have
\[\begin{vmatrix}
-i \omega & ik & 0\\ 
0 & -i \omega & ik \\
1 & 0 & -(k^2 + u_{ph} a_1 \sigma_1 +u_{e} b_1 \sigma_2)
\end{vmatrix} =0 \] 
\begin{equation}\label{dis}
\begin{split}	
 \implies \omega^2 &= \frac{k^2}{(k^2 + u_{ph} a_1 \sigma_1 +u_{e} b_1 \sigma_2)}  \\ 
\implies  \omega^2 &= \frac{k^2}{k^2 + u_{ph} (\frac{\kappa_p - \frac{1}{2}}{\kappa_p - \frac{3}{2}}) \sigma_1 +u_{e} (\frac{\kappa_e-\frac{1}{2}}{\kappa_e -\frac{3}{2}}) \sigma_2},
\end{split}
\end{equation} 
this is the dispersion relation.
	
\section{Derivation of evolution equations}\label{evo}

\subsection{Korteweg-de Vries (KdV) equation}
The KdV equation has been introduced by the following stretched coordinates: \begin{equation}\label{s1}
\xi=\epsilon^{\frac{1}{2}}(x-\lambda t), \tau= \epsilon^{\frac{3}{2}} t,
\end{equation}
Then the derivatives operators are considered as follows: \[\frac{\partial}{\partial x}=\epsilon^{\frac{1}{2}}\frac{\partial}{\partial \xi}, \frac{\partial}{\partial t}= \epsilon^{\frac{3}{2}}\frac{\partial}{\partial \tau}-\lambda \epsilon^{\frac{1}{2}} \frac{\partial}{\partial \xi}, \frac{\partial^2}{\partial x^2}=\epsilon \frac{\partial^2}{\partial \xi^2} . \]
Let the dependent variables are perturbed in the following way: 
\begin{equation}\label{p1}
	n_{pc}=1+\epsilon n^{(1)}_{pc}+\epsilon^2 n^{(2)}_{pc}+\epsilon^3 n^{(3)}_{pc}+ \dots,  
\end{equation}       
\begin{equation}\label{p2}
	u_{pc}=0+\epsilon u^{(1)}_{pc}+\epsilon^2 u^{(2)}_{pc}+ \epsilon^3 u^{(3)}_{pc}+ \dots, 
\end{equation}    
\begin{equation}\label{p3} 
	\phi =0+ \epsilon \phi^{(1)}+ \epsilon^2 \phi^{(2)}+ \epsilon^3 \phi^{(3)}+ \dots .
\end{equation}
First, we express the Eqs. \eqref{n1}, \eqref{n2} and \eqref{n4} in terms of $\xi$ and $\tau$, then substitute the above expression into Eqs. \eqref{n1}, \eqref{n2} and \eqref{n4} . Now equating the coefficients of each power of $\epsilon$, we get the phase velocity 
\begin{equation}\label{pv2_kdV}
\lambda= \left[ \frac{1}{\mu_{ph}a_1 \sigma_1+ \mu_e b_1 \sigma_2} \right]^{1/2}.
\end{equation}
After some straightforward calculations, we obtained 
\begin{equation}\label{KdV2}
\frac{\partial \phi^{(1)}}{\partial \tau}+ A_1 \phi^{(1)} \frac{\partial \phi^{(1)}}{\partial \xi}+ B_1 \frac{\partial^3 \phi^{(1)}}{\partial \xi^3}=0,
\end{equation}
where $ A_1 =  \frac{1}{2(\mu_{ph}a_1 \sigma_1+ \mu_e b_1 \sigma_2)^{3/2}} [ 3(\mu_{ph}a_1 \sigma_1  + \mu_e b_1 \sigma_2)^2 -(\mu_e b_2 \sigma^2_2 -\mu_{ph}a_2 \sigma^2_1)], $
and $ B_1= \frac{\lambda^3}{2}= \frac{1}{2}\left[ \frac{1}{(\mu_{ph}a_1 \sigma_1+ \mu_e b_1 \sigma_2)^{3/2}} \right] .$

\subsection{Modified Korteweg-de Vries (MKdV) equation}
The coefficient of the quadratic nonlinear term ($ A_1 \phi^{(1)} \frac{\partial \phi^{(1)}}{\partial \xi}$) in KdV equation i.e. $A_1$ depends on different parameters, and for a particular parameter set the value of $A_1$ will be $0$, then the equation will become a linear one. So, to consider nonlinearity in the evolution equation a separate stretching is needed and accordingly, a new higher-order term i.e. cubic nonlinear term appears in the evolution equation. 
For this reason, we employ the following stretched coordinates for mKdV equation: \[ \xi= \epsilon(x-\lambda t), \tau= \epsilon^3 t. \] 
Now considering the above perturbed variables (\ref{p1})-(\ref{p3}), and equating the lowest power of $\epsilon$ to zero, we obtain the phase velocity as
\begin{equation}\label{pv2_mkdv}
\lambda=  \left[ \frac{1}{(\mu_{ph}a_1 \sigma_1+\mu_e b_1 \sigma_2)} \right]^{1/2}.
\end{equation}  
Then considering the order of $\epsilon^3$, we obtain
\begin{equation}\label{A0}
-\frac{1}{2}A_1 (\phi^{(1)})^2 = 0\implies A_1=0 .
\end{equation} 
After some calculations, we obtain the mKdV equation in terms of variable $\phi^{(1)}$, 
\begin{equation}\label{mKdV2}
\frac{\partial \phi^{(1)}}{\partial \tau}+B_2 \frac{\partial^3 \phi^{(1)}}{\partial \xi^3}+B_2 C_2 (\phi^{(1)})^2 \frac{\partial \phi^{(1)}}{\partial \xi} =0,
\end{equation}
where \[B_2=\frac{\lambda^3}{2}=(B_1), C_2= \frac{15}{2 \lambda^6}- (\mu_{ph}\frac{a_3}{2}\sigma^3_1 + \mu_e \frac{b_3}{2} \sigma^3_2). \]

\subsection{Gardner equation (GE)}
The derivation and analysis of the KdV and mKdV equations lay the groundwork for comprehending how nonlinearity and dispersion interact during wave propagation. Building on this foundational research, the GE investigates a wider range of nonlinear phenomena, such as breathers and solitons. So, in this section, we will derive the GE, and then we will discuss the various nonlinear structures of the GE.\\
For $\mu_{ph}$ around its critical value $(\mu_c)$, $A_1=A_0$ can be expressed as: \[ A_0 \approx s \left( \frac{\partial A_1}{\partial \mu_{ph}}\right)_{\mu_{ph}=\mu_c}|\mu_{ph}-\mu_c| = c_1 s \epsilon , \]
where $|\mu_{ph}-\mu_c|$ is a small and dimensionless parameter and can be taken as the expansion parameter $\epsilon$,   i.e., $|\mu_{ph}-\mu_c| \approx \epsilon $, and $s=1$ for $\mu_{ph}>\mu_c $. $s=-1$ for $\mu_{ph}<\mu_c $. $c_1 = \frac{\partial A_1}{\partial \mu_{ph}} $ is a constant depending on the plasma parameters $\sigma_1, \sigma_2, \mu_{ph}, \mu_e, \kappa_e, \kappa_p. $ Now considering the third order of $\epsilon$ in the Poisson's equation, we obtain the Gardner equation (replacing $\phi^{(1)}$ by $\phi$) as: 
\begin{equation}\label{GE2}
\frac{\partial \phi}{\partial \tau}+ A_3 \phi \frac{\partial \phi} {\partial \xi}+ C_3 (\phi)^2 \frac{\partial \phi}{\partial \xi}+B_3 \frac{\partial^3 \phi}{\partial \xi^3}=0,
\end{equation} 
where $A_3=c_1 s B_2, C_3=B_2 C_2, B_3=(B_2=B_1).$ 
The Gardner equation is also called the mixed KdV equation. It contains both $\phi$ term of KdV and $(\phi)^2$ term of mKdV equation. 
	
\section{Analytical solutions of Gardner equation}\label{analytic}
To find the soliton solution of the GE \eqref{GE2}, we take the transformation 
\begin{equation}\label{T}
\phi=2[tan^{-1}(g/f)]_{\xi}.
\end{equation}
Then the Hirota bilinear forms of \eqref{GE2} are \begin{equation}\label{HB1}
	(D_{\tau}+ B_3 D^3_{\xi})(g.f)=0,
\end{equation} and 
\begin{equation} \label{HB2}
	3 B_3 D^2_{\xi}(f.f+g.g)- A_3 D_{\xi}(g.f)=0 . 
\end{equation}
	
\subsection{One-soliton solution} \label{1sol}
For one-soliton solution  we take \[ f=1+exp(\theta), \] and \[ g=1+d\; exp(\theta), \] where $\theta= p~ \xi- \Omega \tau-\xi_0 $; $p$ is the wave number and $\Omega$ is the frequency and $\xi_0$ is the initial phase.	
Using the values of $f$ and $g$ into the Eqs. \eqref{HB1} and \eqref{HB2}, we get the dispersion relation $ \Omega= B_3 p^3 $ and $d=\frac{A_3 + 6 B_3 p}{A_3 - 6 B_3 p}.$ \\

For $ A_3 > C_3~ p$, the one-soliton solution is,
\begin{equation}\label{1-S_1GE}
	\phi(\xi,\tau)= \frac{ C_3 p^2}{A_3 + \sqrt{(A_3)^2 + ( C_3)^2 p^2} cosh [p(\xi -B_3 p^2 \tau - \xi_0)]},
\end{equation} 
where \[ exp(p \xi_0)=\frac{A_3-  C_3 p}{\sqrt{(A_3)^2 + ( C_3)^2 p^2}}, \] and hence the amplitude is \[ \frac{ C_3 p^2}{A_3 + \sqrt{(A_3)^2+ ( C_3)^2 p^2}}. \]
\begin{figure}[H]
\centering
\subfigure[h]{\label{1a_1solitoncase1_2d1}
\includegraphics[width=0.4\linewidth]{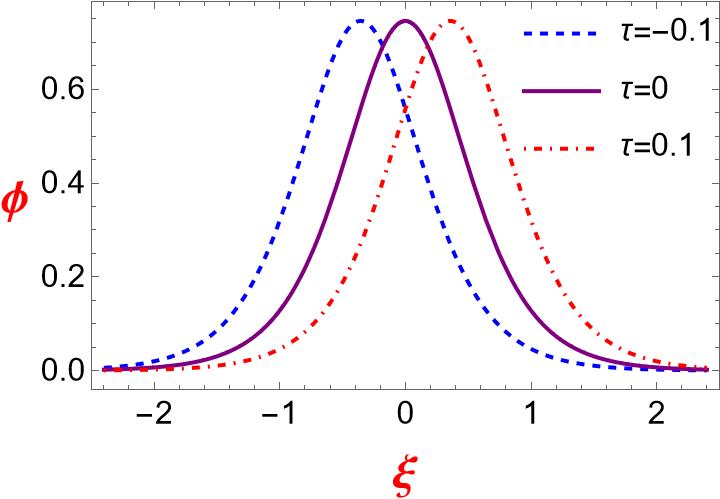}} 
\hspace{0.2 in}
\subfigure[h]{\includegraphics[width=0.42\linewidth]{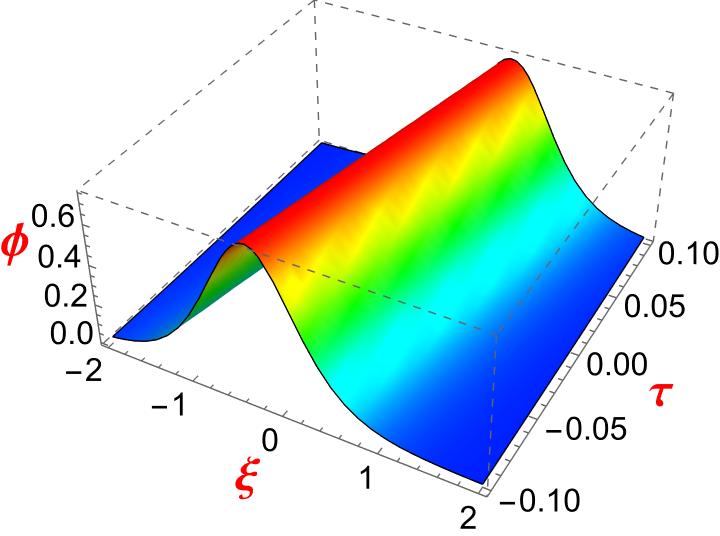}
\label{1b_1solitoncase1_3d1}}
\caption{(a) The 2D graph and (b) The 3D graph of one-soliton solution, by considering $B_3(=B_2)=0.396821, C_2=0.457408, c_1=2.5902, s=  1, p= 3$.}
\end{figure}
\begin{figure}[H]
\centering
\subfigure[]{\includegraphics[width=0.42\linewidth]{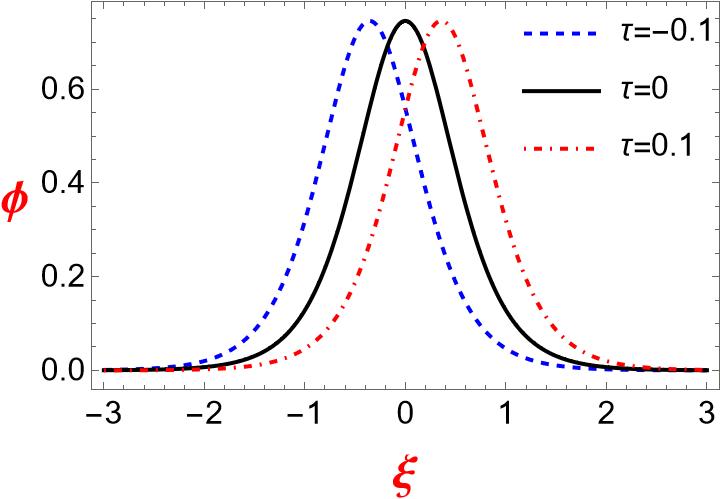} \label{2a_1solitoncase1_2d2}}
\hspace{0.2 in}
\subfigure[]{\includegraphics[width=0.42\linewidth]{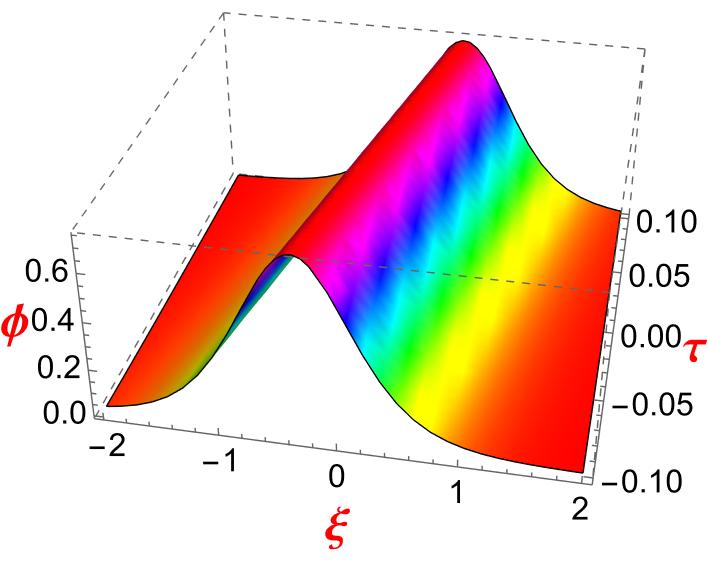}\label{2b_1solitoncase1_3d2}}	 
\caption{(a) The 2D graph and (b) The 3D graph of one-soliton solution, when $B_3(=B_2)=0.396821, C_2=0.457408, c_1=2.5902, s=1, p=-3$.}
\end{figure}

The one-soliton solution of Eq. \eqref{GE2} occurs
for $A_3 <  C_3~ p $ and given by  
\begin{equation}\label{1-S_2GE}
	\phi(\xi, \tau)= - \frac{ C_3 p^2}{\sqrt{(A_3)^2 + ( C_3)^2 p^2} cosh [p(\xi - B_3 p^2 \tau- \xi_0)]-A_3},
\end{equation} 
where \[ exp(p \xi_0)= \frac{ C_3 p - A_3}{\sqrt{(A_3)^2 + ( C_3)^2 p^2}}. \]
\begin{figure}[H]
\centering
\subfigure[]{\includegraphics[width=0.4\linewidth]{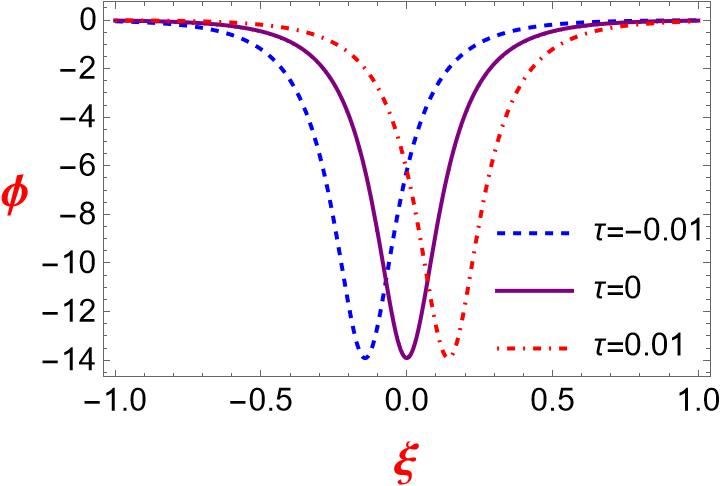}\label{3a_1solitoncase2_2d} }\hspace{0.2 in}
\subfigure[]{\includegraphics[width=0.42\linewidth]{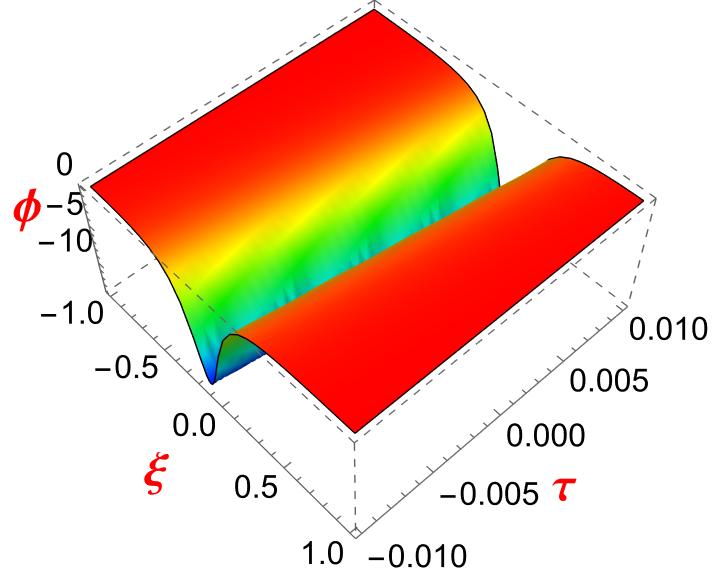}\label{3b_1solitoncase2_3d}}	 
\caption{(a) The 2D graph and (b) The 3D graph of one-soliton solution, by considering $B_3(=B_2)=0.396821, C_2=0.457408, c_1=2.5902, s=1, p=6$.}
\end{figure}

\subsection{Two-soliton solutions}\label{2sol}
For this we take \[ f= 1+ exp(\theta)+ exp(\psi)+ f_{12}exp(\theta+\psi), \] and \[ g= 1+ d_1 exp(\theta)+ d_2 exp(\psi)+ d_{12}exp(\theta+ \psi),\] where $\theta=p_1 \xi-\Omega_1 \tau, $ and $\psi=p_2 \xi-\Omega_2 \tau;$ $p_1, p_2$ are the wave numbers and  $\Omega_1,\Omega_2 $ are the frequencies. Using the values of $f$ and $g$ from Eqs. \eqref{HB1} and \eqref{HB2}, we get the dispersion relation,  
\[ \Omega_i=B_3 p_i^3 , ~i= 1,2. \]  
Now the 2-soliton solutions of equ.\eqref{GE2} is given by 
\begin{equation}\label{2ss} 
	\phi=[2 tan^{-1}(g/f)]_{\xi},
\end{equation} where
\[ \theta= p_1 \xi- B_3 p_1^3 \tau, \psi= p_2 \xi- B_3 p_2^3 \tau, d_n= \frac{A_3+  C_3 p_n}{A_3-  C_3 p_n}, n=1,2, \] and \[ f_{12}=\frac{(p_1 -p_2)^2[(A_3)^2 -(A_3) ( C_3)(p_1 + p_2)-(C_3)^2 p_1 p_2]}{(p_1 + p_2)^2 ( A_3 - C_3 p_1)( A_3 -   C_3 p_2)}, \] \[d_{12}=\frac{(p_1 -p_2)^2 [ (A_3)^2 + (A_3) ( C_3) (p_1+p_2)-( C_3)^2 p_1 p_2]}{(p_1+p_2)^2 (A_3- C_3 p_1)(A_3-  C_3 p_2)} \].
\begin{figure}[H]
\centering
\subfigure[]{\includegraphics[width=0.42\linewidth]{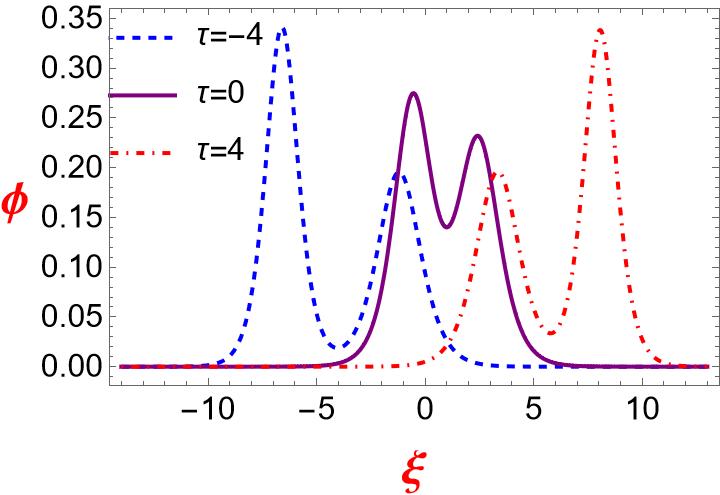}\label{4a_2soliton_2d1}} \hspace{0.2 in}
\subfigure[]{\includegraphics[width=0.42\linewidth]{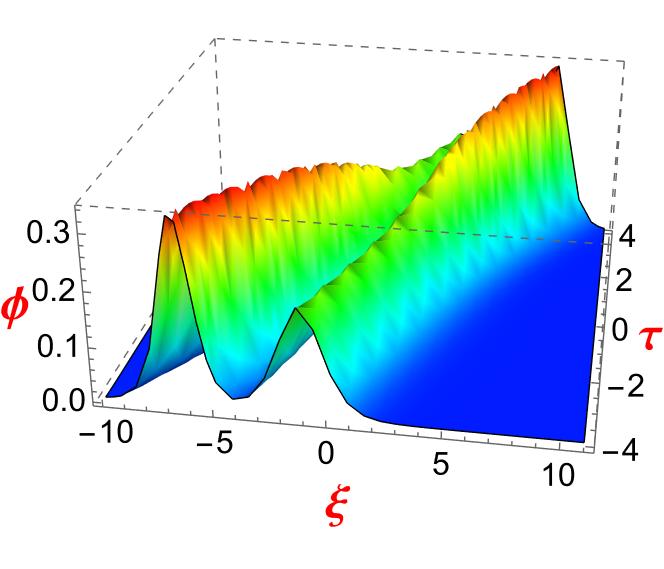}\label{4b_2soliton_3d1} }
\caption{(a) The 2D graph and (b) The 3D graph of two-soliton solutions, by considering $B_3(=B_2)=0.396821, C_2=0.457408, c_1=2.5902, p_1=2, p_2=1.5, s=1$.}
\end{figure}
\begin{figure}[H]
\centering
\subfigure[]{\includegraphics[width=0.42\linewidth]{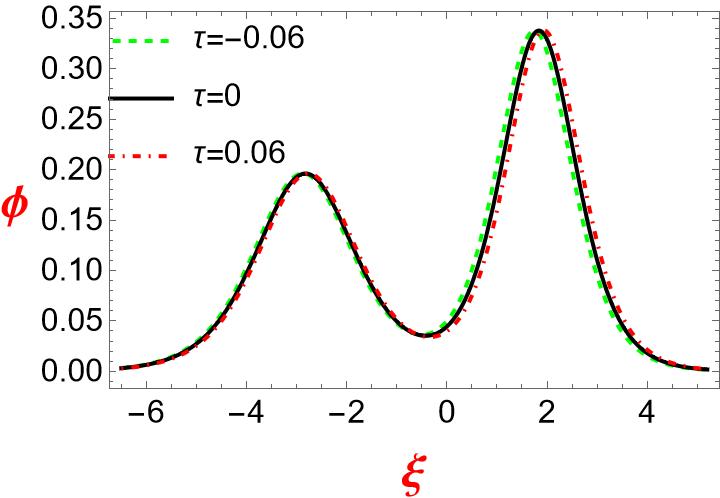}\label{5a_2soliton_2d2}} \hspace{0.2 in}
\subfigure[]{\includegraphics[width=0.42\linewidth]{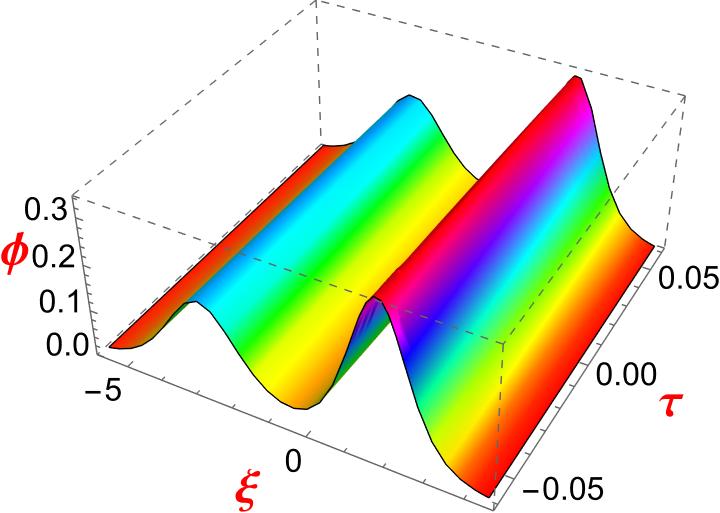}\label{5b_2soliton_3d2} }
\caption{(a) The 2D graph and (b) The 3D graph of two-soliton solutions, by considering $B_3(=B_2)=0.396821, C_2=0.457408, c_1=2.5902, p_1=-2, p_2=1.5, s=1$.}
\end{figure}
\begin{figure}[H]
\centering
\subfigure[]{\includegraphics[width=0.42\linewidth]{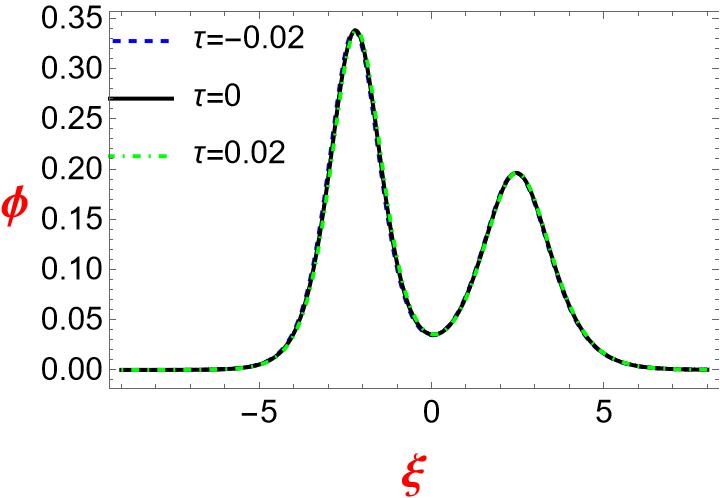} \label{6a_2soliton_2d3}} \hspace{0.2 in}
\subfigure[]{\includegraphics[width=0.42\linewidth]{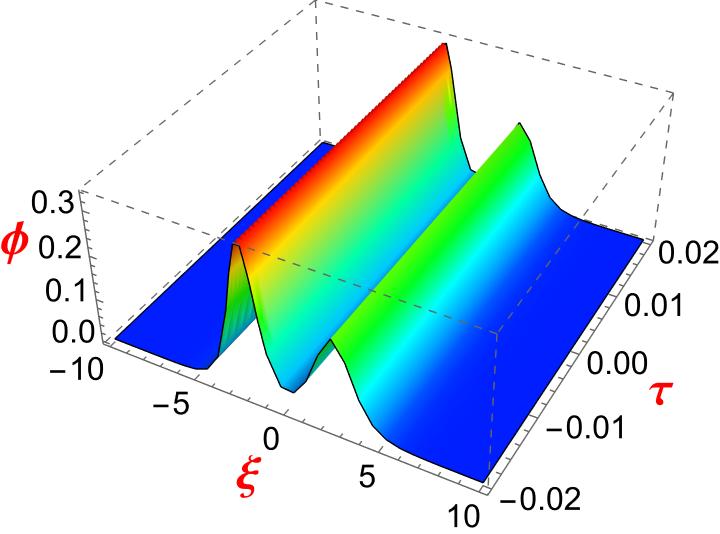} \label{6b_2soliton_3d3} }
\caption{(a) The 2D graph and (b) The 3D graph of two-soliton solutions, by considering $B_3(=B_2)=0.396821, C_2=0.457408, c_1=2.5902, p_1=2, p_2=-1.5, s=1$.}
\end{figure}
\begin{figure}[H]
\centering
\subfigure[]{\includegraphics[width=0.42\linewidth]{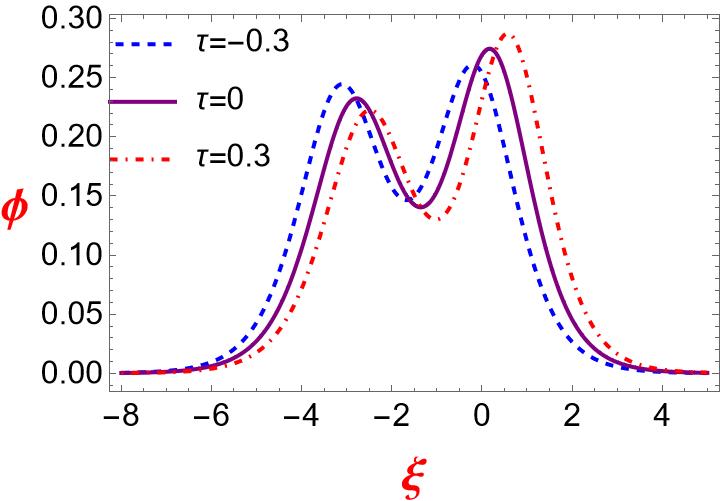} \label{7a_2soliton_2d4} }\hspace{0.2 in}
\subfigure[]{\includegraphics[width=0.42\linewidth]{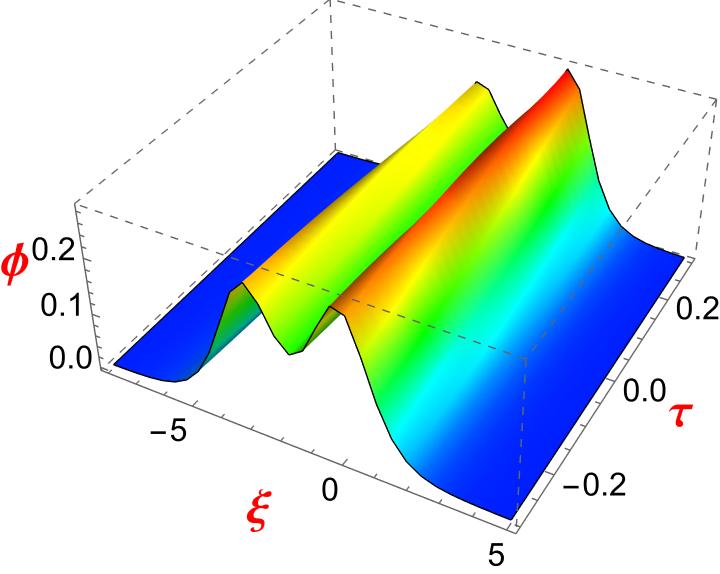} \label{7b_2soliton_3d4}} 
\caption{(a) The 2D graph and (b) The 3D graph of two-soliton solutions, by considering $B_3(=B_2)=0.396821, C_2=0.457408, c_1=2.5902, p_1=-2, p_2=-1.5, s=1$.}
\end{figure}

\subsection{Breather Solutions}\label{breathersol}
The breather is obtained from two-soliton solutions by choosing a pair of complex conjugate wave numbers ($ p_1=m+in, p_2=m-in,$ $m,n$ are real). After some calculations, we obtain the breather solution of GE, given as below: 
\begin{equation}\label{bsGE}
\phi(\xi, \tau)=[2 tan^{-1}(g/f)]_{\xi},
\end{equation}
where \begin{multline*}
g=1- \frac{n^2}{m^2}\left[  \frac{(A_3)^2 +2 (A_3) ( C_3) m - ( C_3)^2 (m^2+n^2)}{(A_3-  C_3 m)^2+ (  C_3)^2 n^2}\right]  exp(2 \Theta) 
+ 2\left[\zeta cos (\Psi) -\eta sin (\Psi)\right] exp(\Theta),
\end{multline*}
\begin{multline*}
f= 1- \frac{n^2}{m^2}\left[ \frac{(A_3)^2 -2 (A_3) (  C_3 ) m -( C_3)^2 (m^2 + n^2)}{(A_3 -  C_3 m)^2+ ( C_3)^2 n^2} \right]  exp( 2 \Theta)  + 2 cos (\Psi) exp(\Theta), 
\end{multline*}
\[ \Theta= m[\xi - B_3 (m^2 -3 n^2)\tau], \Psi= n[\xi - B_3 (3 m^2 -n^2)\tau], \] and \[ \zeta = \frac{( A_3)^2 -( C_3)^2 (m^2 + n^2)}{(A_3 -  C_3 m)^2+ ( C_3)^2 n^2} , \eta= \frac{ 2 (A_3) (  C_3) n }{(A_3 -  C_3 m)^2 + ( C_3)^2 n^2}. \] 
	
\begin{figure}[H]
\centering
\subfigure[]{\includegraphics[width=0.42\linewidth]{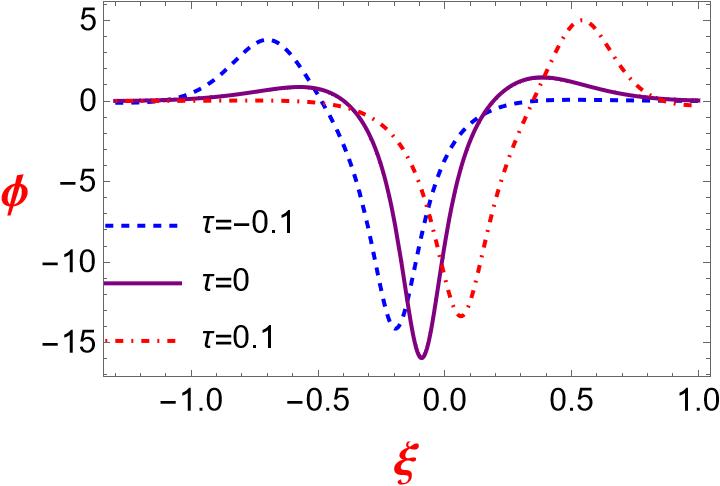}\label{8a_breather_2d2}} \hspace{0.2 in}
\subfigure[]{\includegraphics[width=0.42\linewidth]{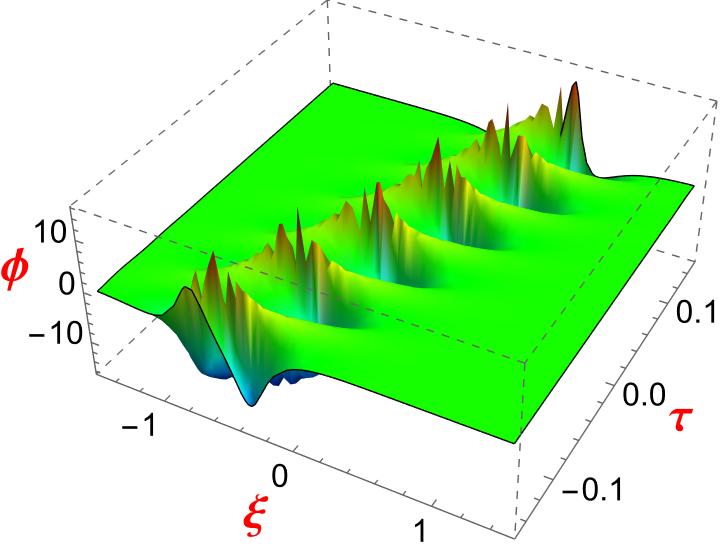} \label{8b_breather_3d2} }
\caption{(a) The 2D graph and (b) The 3D graph of breather solution, by considering $B_3(=B_2)=0.396821, C_2=0.457408, c_1=2.5902,  m=6, n=3, s= 1$.}
\end{figure}
\begin{figure}[H]
\centering
\subfigure[]{\includegraphics[width=0.43\linewidth]{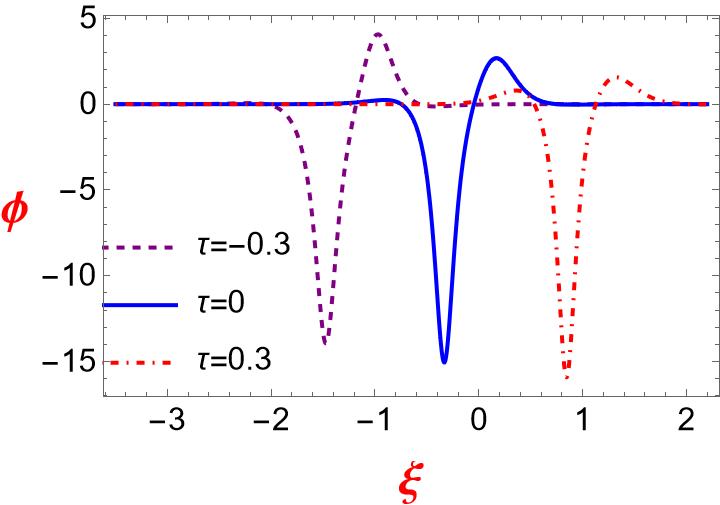}\label{9a_breather_2d2}} \hspace{0.2 in}
\subfigure[]{\includegraphics[width=0.42\linewidth]{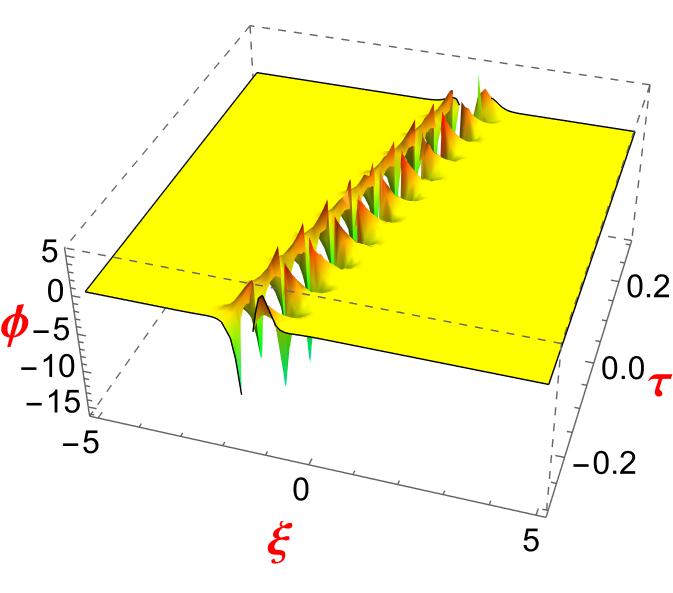} \label{9b_breather_3d2} }
\caption{(a) The 2D graph and (b) The 3D graph of breather solution, by considering $B_3(=B_2)=0.396821,C_2=0.457408,c_1=2.5902,m=-6,n=-3,s= 1$.}
\end{figure}
\begin{figure}[H]
\centering
\subfigure[]{\includegraphics[width=0.42\linewidth]{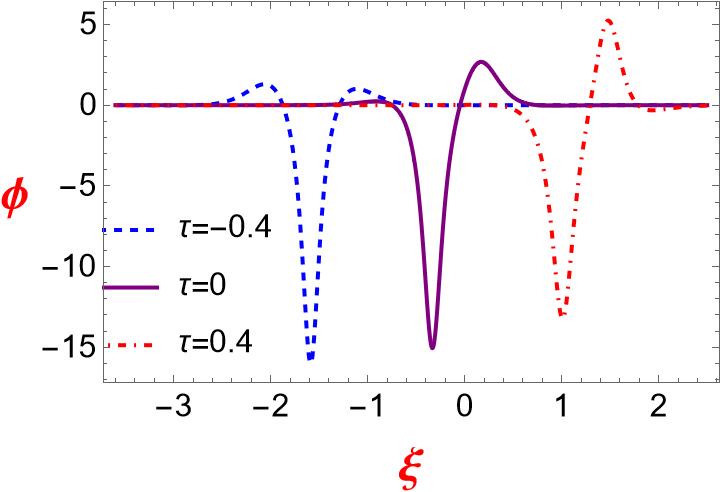} \label{10a_breather_2d2}} \hspace{0.1 in}
\subfigure[]{\includegraphics[width=0.46\linewidth]{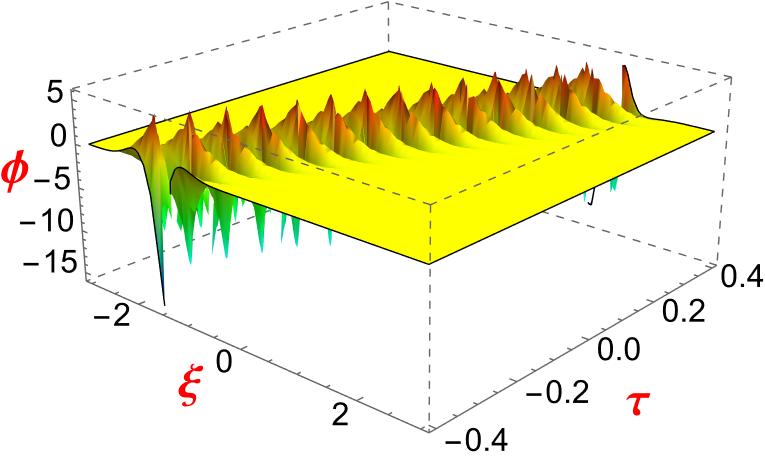} \label{10b_breather_3d2} }
\caption{(a) The 2D graph and (b) The 3D graph of breather solution, by considering $B_3(=B_2)=0.396821, C_2=0.457408, c_1=2.5902,  m=-6, n=3, s= 1$.}
\end{figure}
\begin{figure}[H]
\centering
\subfigure[]{\includegraphics[width=0.44\linewidth]{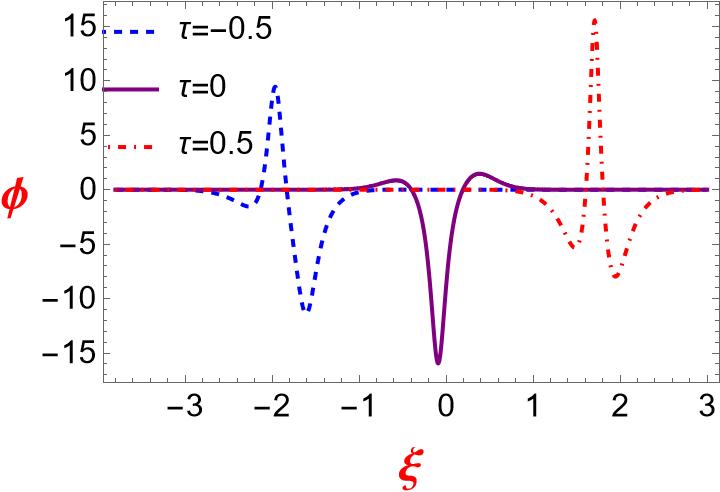} \label{11a_breather_2d2}} \hspace{0.2 in}
\subfigure[]{\includegraphics[width=0.42\linewidth]{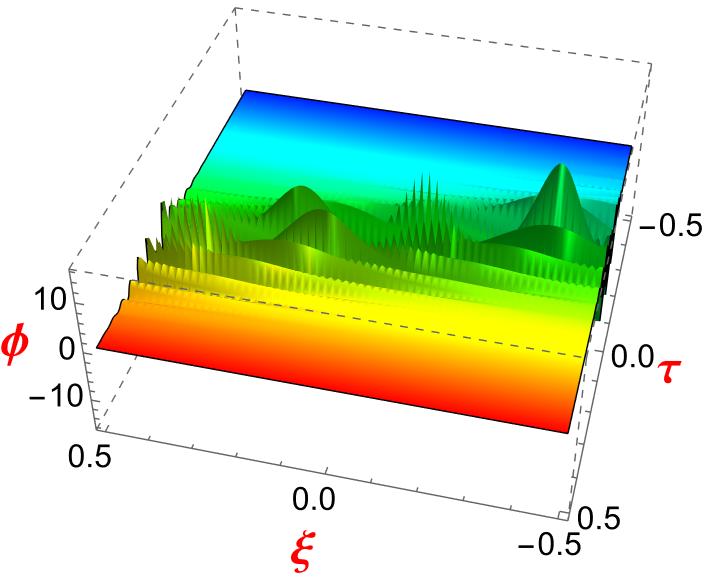} \label{11b_breather_3d2} }
\caption{(a) The 2D graph and (b) The 3D graph of breather solution, by considering $B_3(=B_2)=0.396821,C_2=0.457408,c_1=2.5902,m=6,n=-3,s= 1$.}
\end{figure}
\begin{figure}[H]
\centering
\subfigure[]{\includegraphics[width=0.42\linewidth]{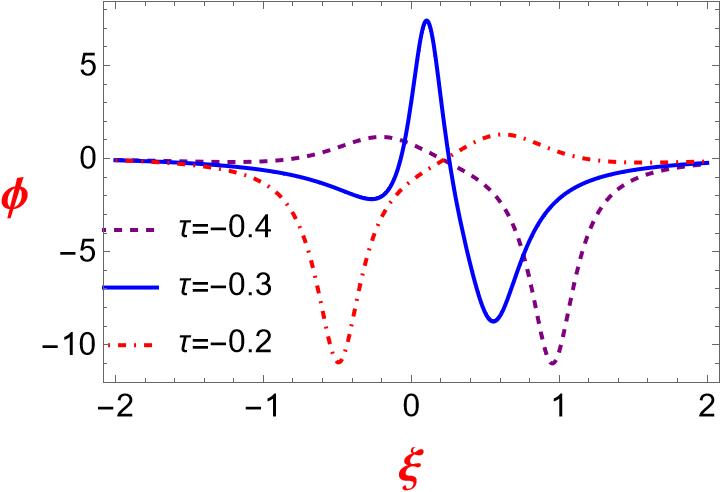}\label{12a_breather_2d1}}\hspace{0.2 in}
\subfigure[]{\includegraphics[width=0.42\linewidth]{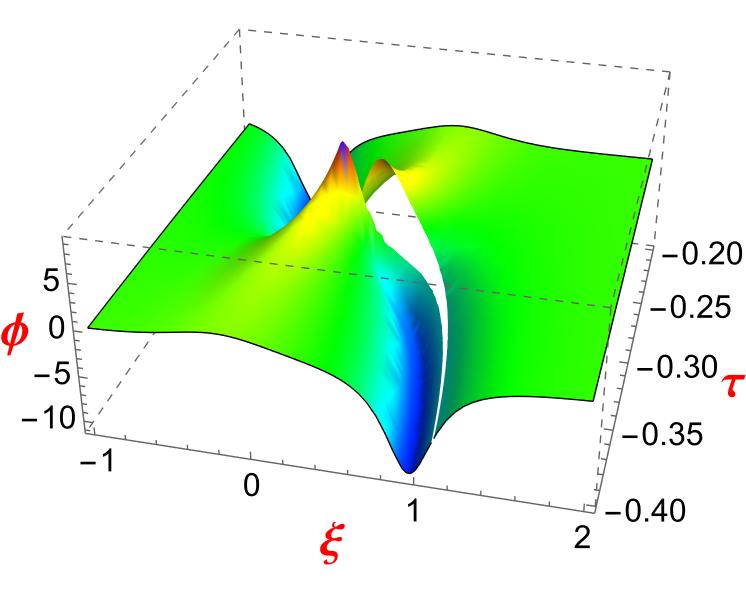} \label{12b_breather_3d1} }
\caption{(a) The 2D graph and (b) The 3D graph of breather solution, by considering $B_3(=B_2)=0.396821,C_2=0.457408,c_1=2.5902,m=-1.7,n=1.5,s= 1$.}
\end{figure}
	
\begin{figure}[H]
\centering
\subfigure[]{\includegraphics[width=0.42\linewidth]{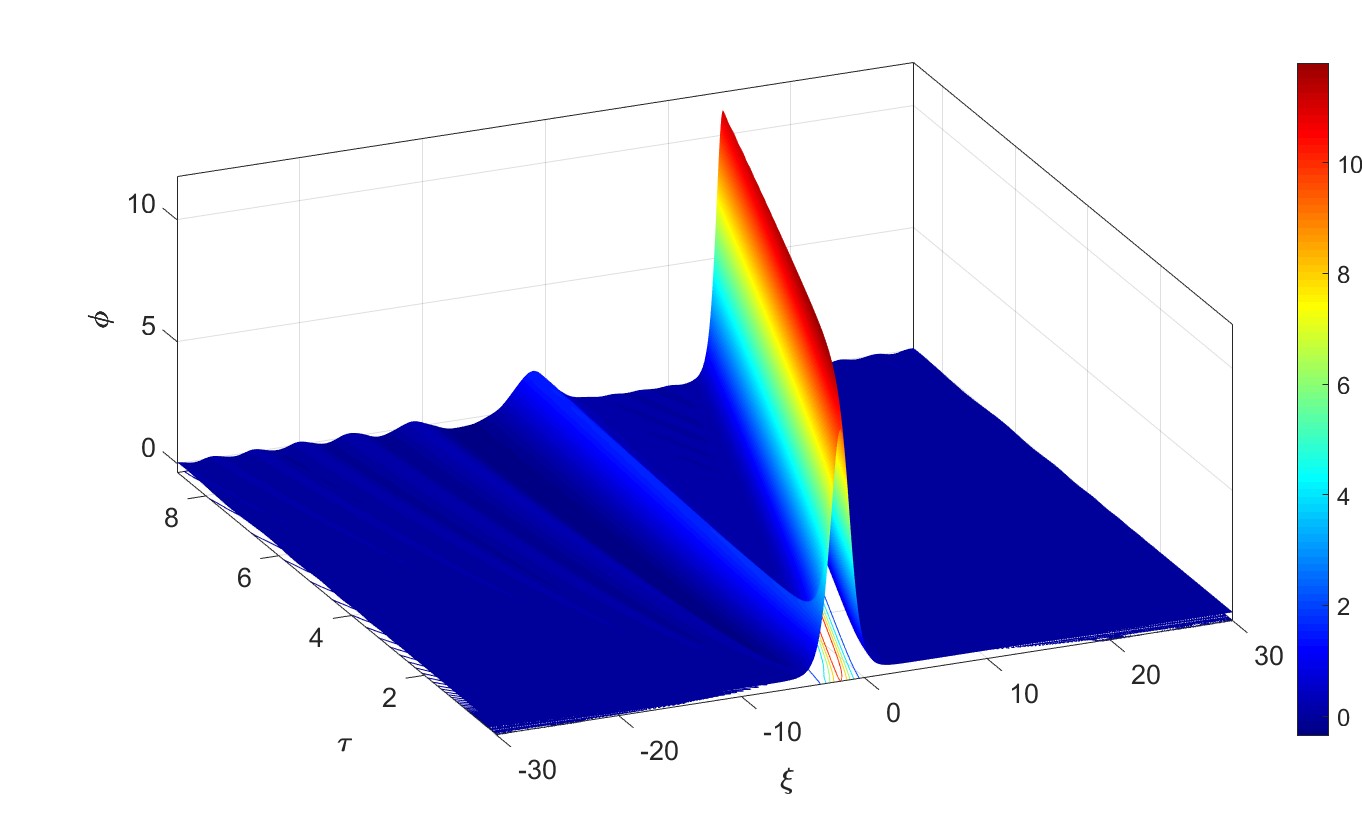}\label{C_kdv_1a}} \hspace{0.2 in}
\subfigure[]{\includegraphics[width=0.42\linewidth]{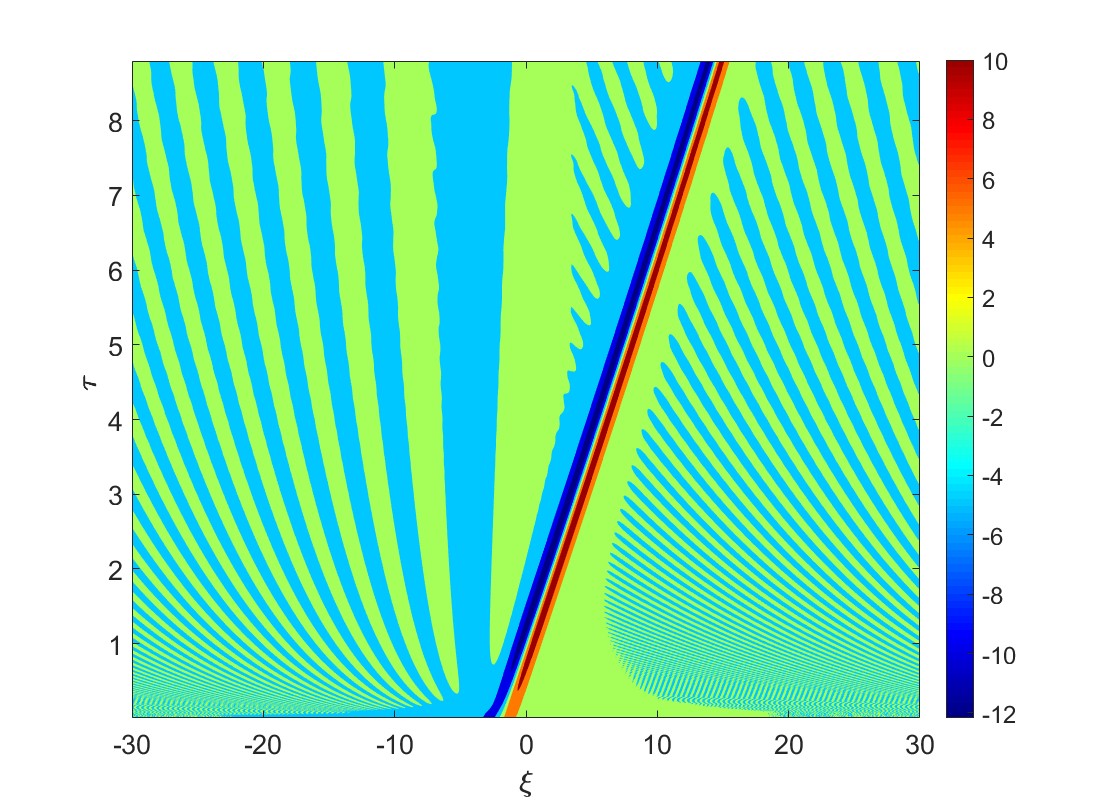}\label{C_kdv_1b}}
\caption{(a) Potential profile and (b) Field contours of numerical solution of KdV equation with positive amplitude single soliton as initial profile, by considering $\mu_e=0.7,\mu_i=-0.4,\mu_ph=0.1,s= 1,p=1$.}
\end{figure}
	
\begin{figure}[H]
\centering
\subfigure[]{\includegraphics[width=0.42\linewidth]{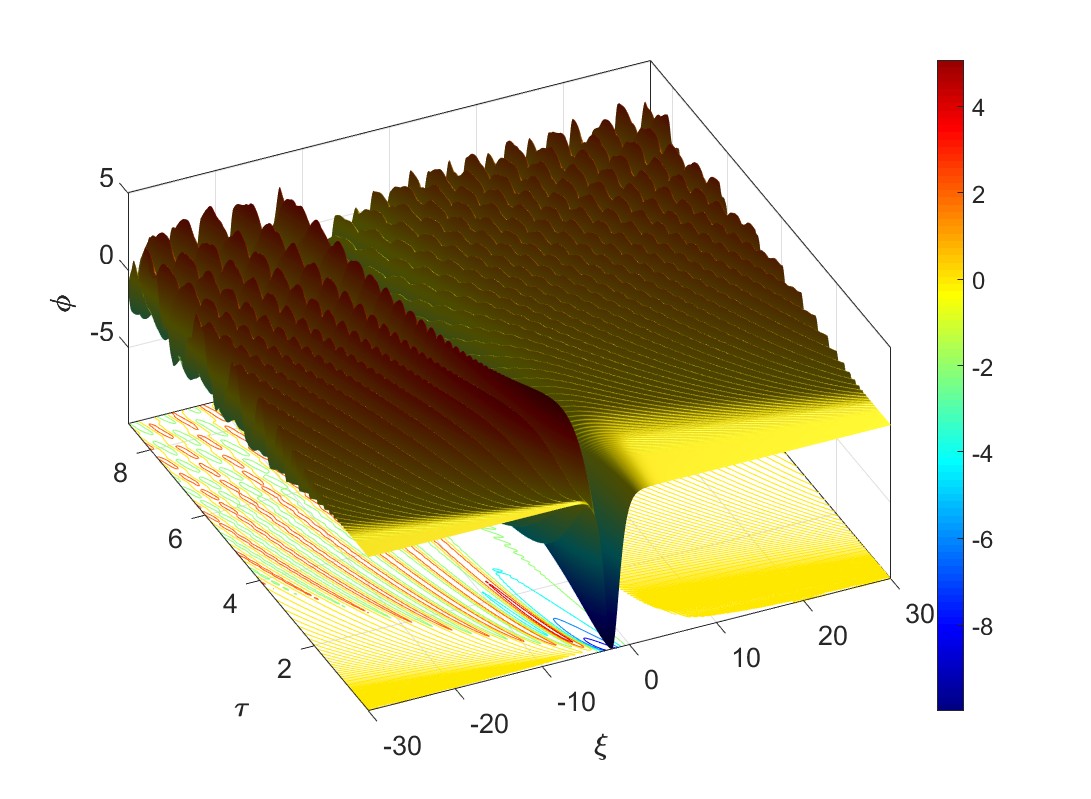}\label{C_kdv_2a}} \hspace{0.2 in}
\subfigure[]{\includegraphics[width=0.42\linewidth]{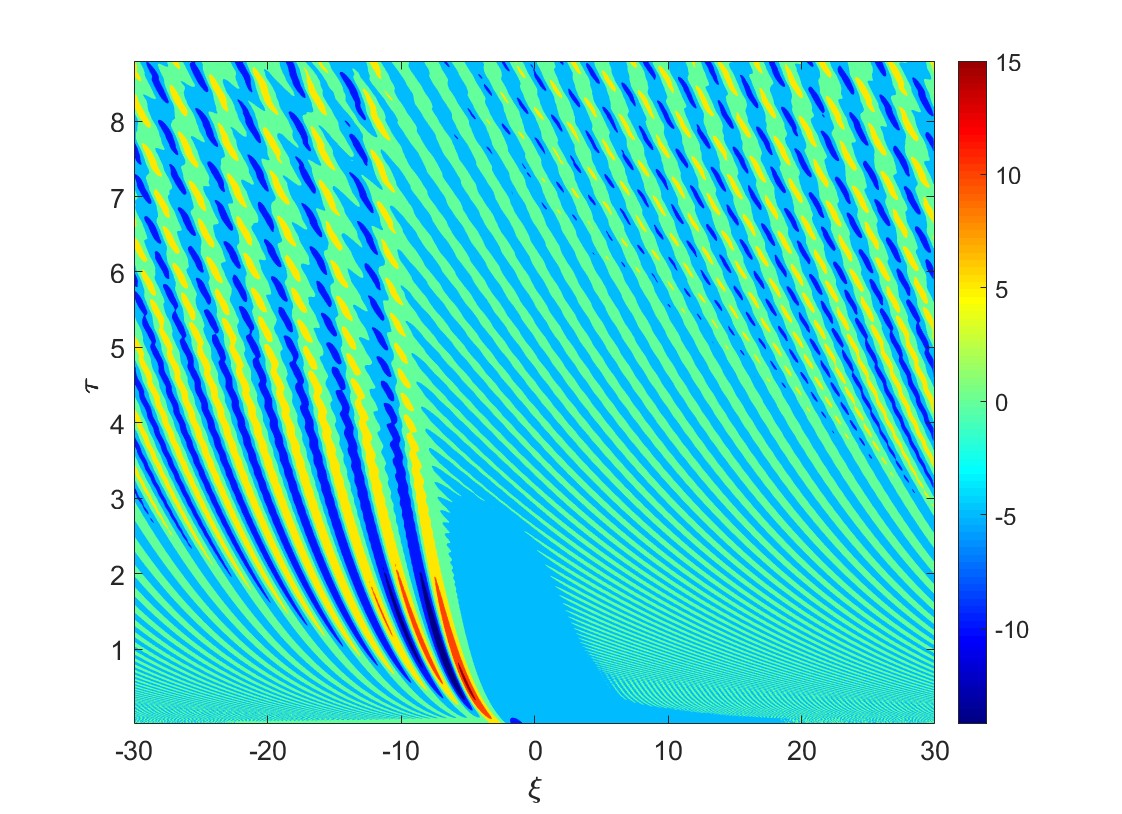}\label{C_kdv_2b}}
\caption{(a) Potential profile and (b) Field contours of numerical solution of KdV equation with negative amplitude single soliton as initial profile, by considering $\mu_e=0.7,\mu_i=-0.4,\mu_ph=0.1,s= 1,p=1$.}
\end{figure}
	
\begin{figure}[H]
\centering
\subfigure[]{\includegraphics[width=0.42\linewidth]{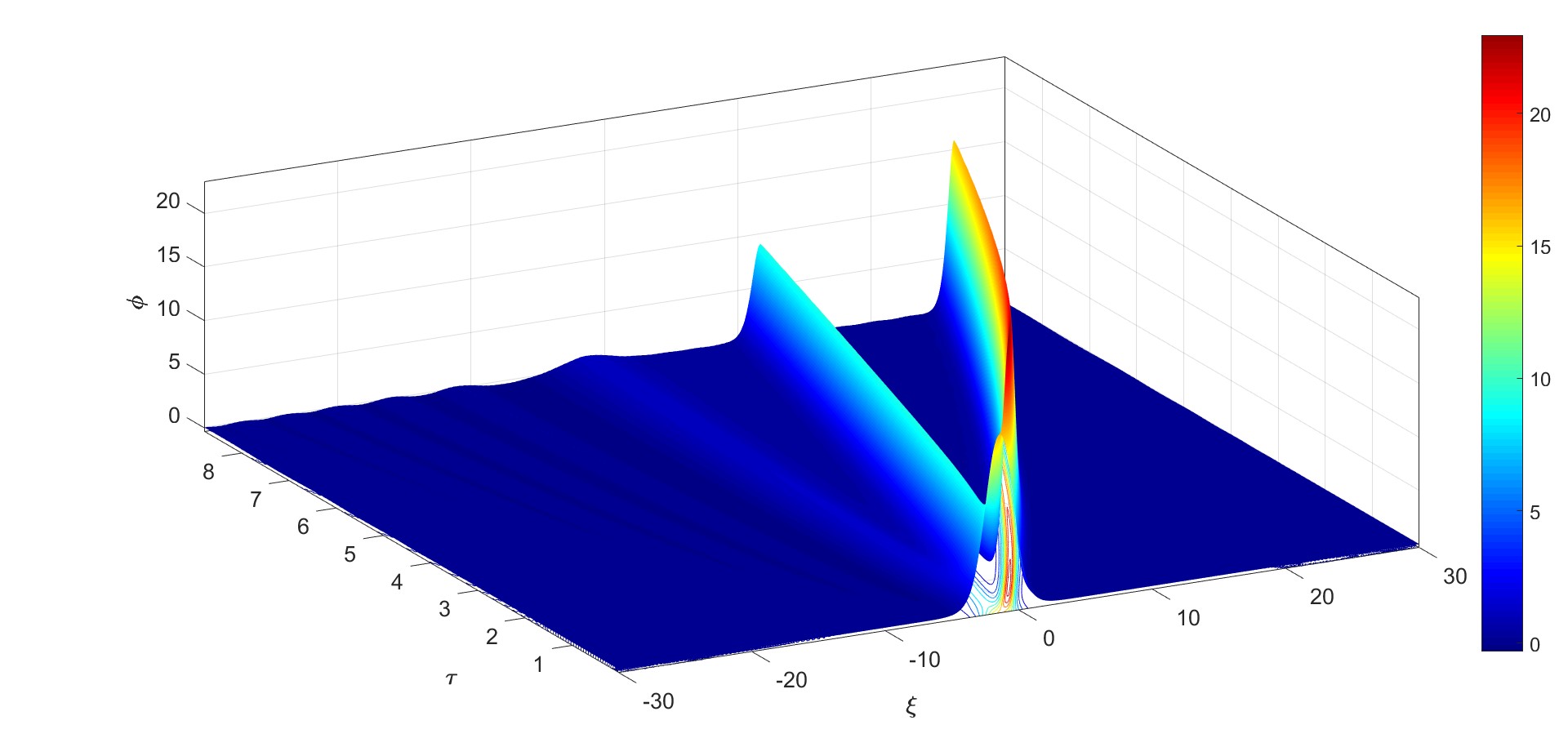}\label{C_kdv_3a}} \hspace{0.2 in}
\subfigure[]{\includegraphics[width=0.42\linewidth]{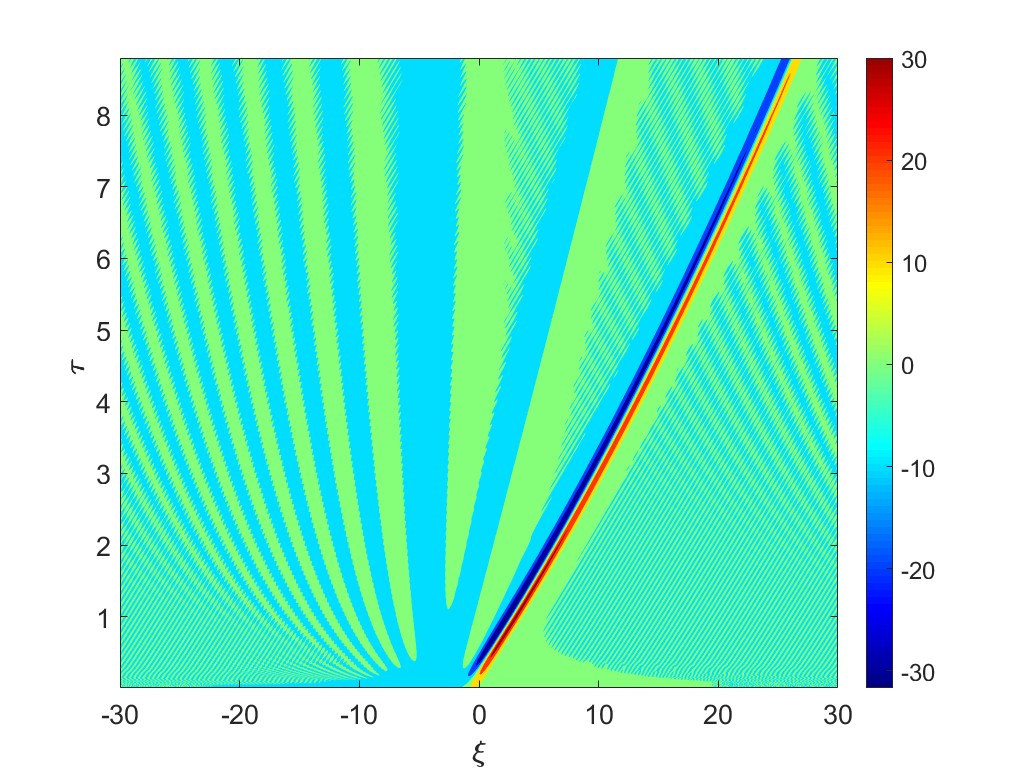}\label{C_kdv_3b}}
\caption{(a) Potential profile and (b) Field contours of numerical solution of KdV equation with positive amplitude double solitons as initial profile, by considering $\mu_e=0.7,\mu_i=-0.4,\mu_ph=0.1,s= 1,p=1$.}
\end{figure}
	
\begin{figure}[H]
\centering
\subfigure[]{\includegraphics[width=0.42\linewidth]{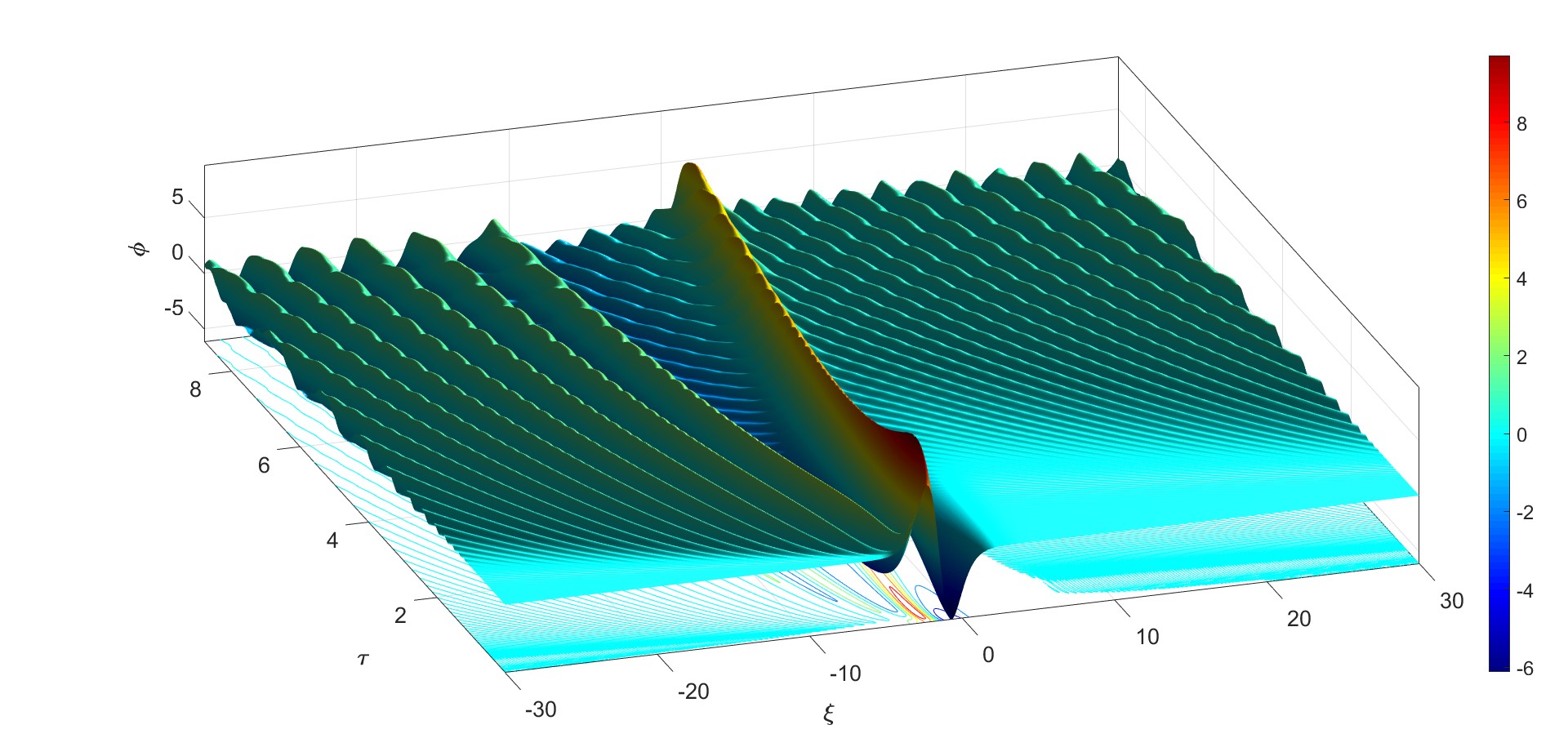}\label{C_kdv_4a}} \hspace{0.2 in}
\subfigure[]{\includegraphics[width=0.42\linewidth]{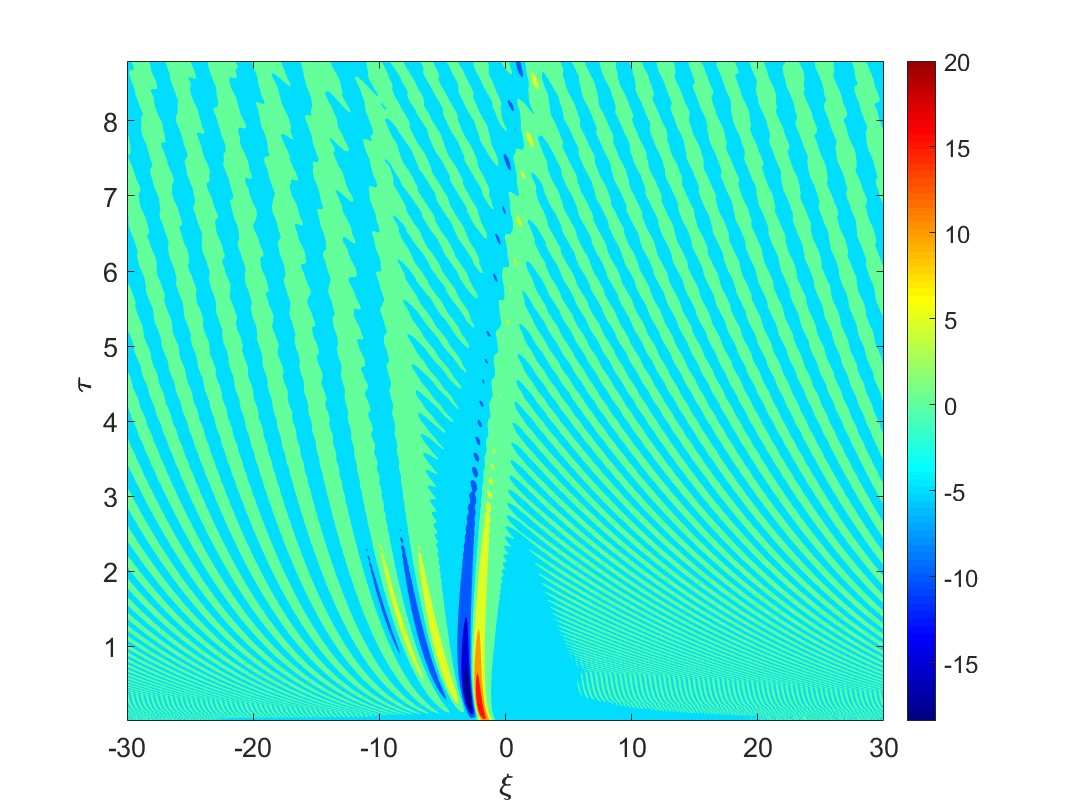}\label{C_kdv_4b}} 
\caption{(a) Potential profile and (b) Field contours of numerical solution of KdV equation with positive and negative amplitude double solitons as initial profile, by considering $\mu_e=0.7,\mu_i=-0.4,\mu_ph=0.1,s= 1,p=1$.}
\end{figure}
	
\begin{figure}[H]
\centering
\subfigure[]{\includegraphics[width=0.42\linewidth]{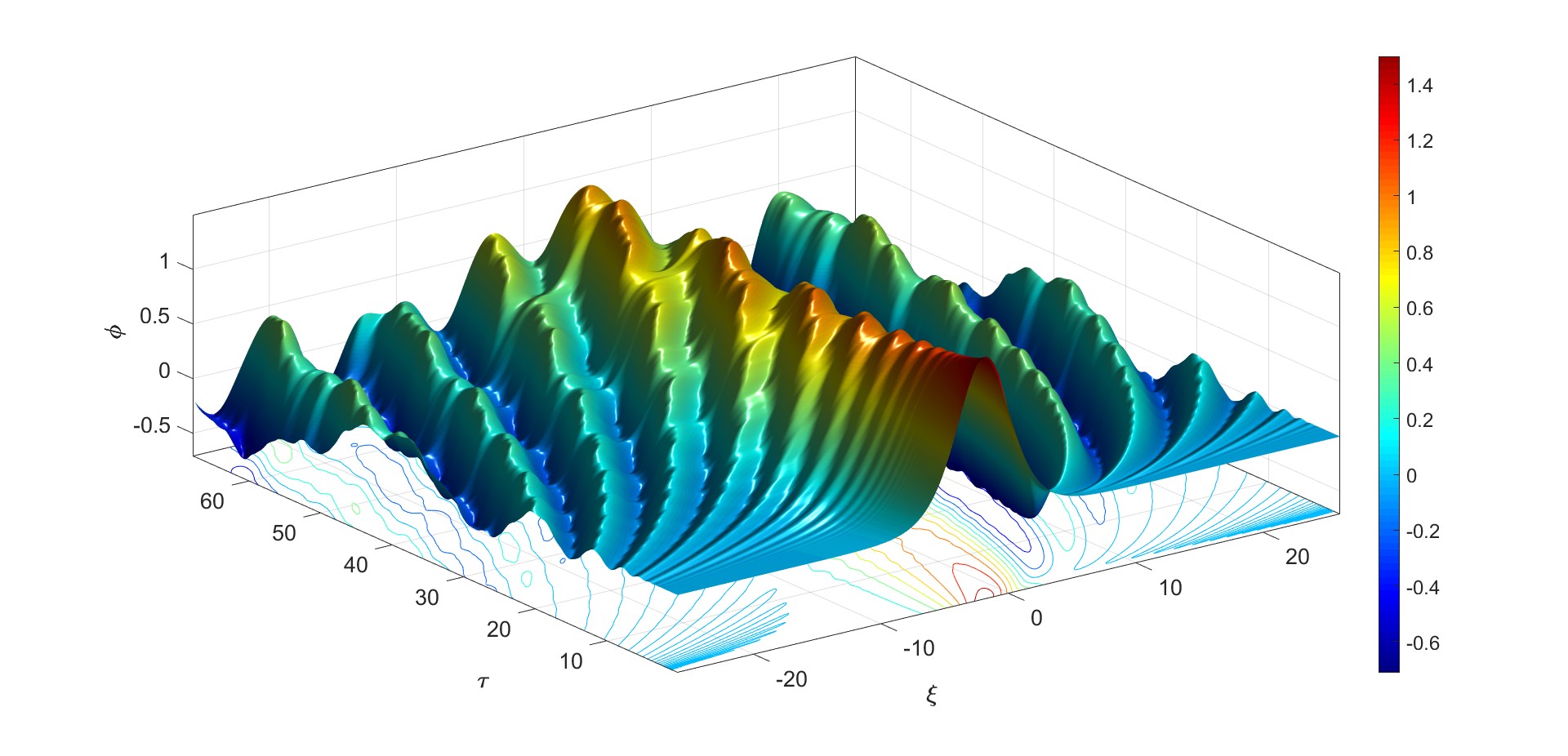} \label{C_mkdv_1a}} \hspace{0.2 in}
\subfigure[]{\includegraphics[width=0.42\linewidth]{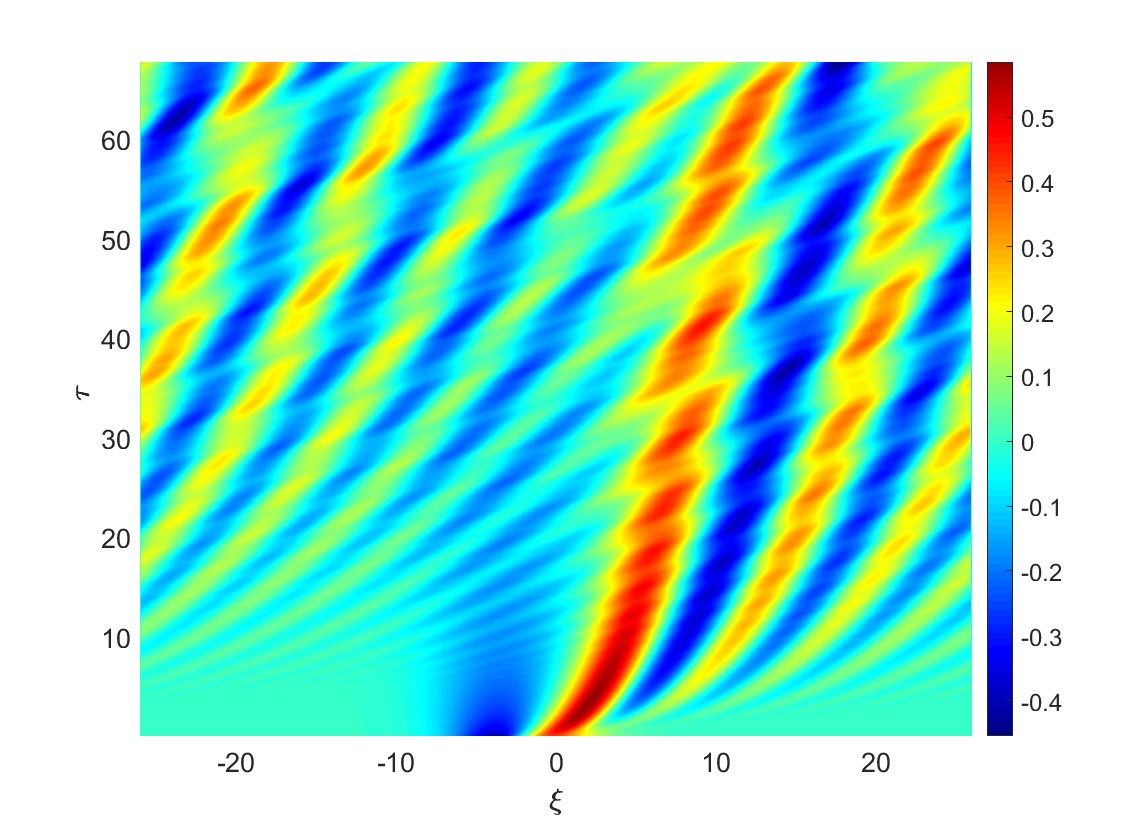} \label{C_mkdv_1b} }
\caption{(a) Potential profile and (b) Field contours of numerical solution of mKdV equation with positive amplitude single soliton as initial profile, by considering $\mu_e=0.7,\mu_i=-0.6,\mu_ph=0.1,s= 1,p=1$.}
\end{figure}
	
\begin{figure}[H]
\centering
\subfigure[]{\includegraphics[width=0.42\linewidth]{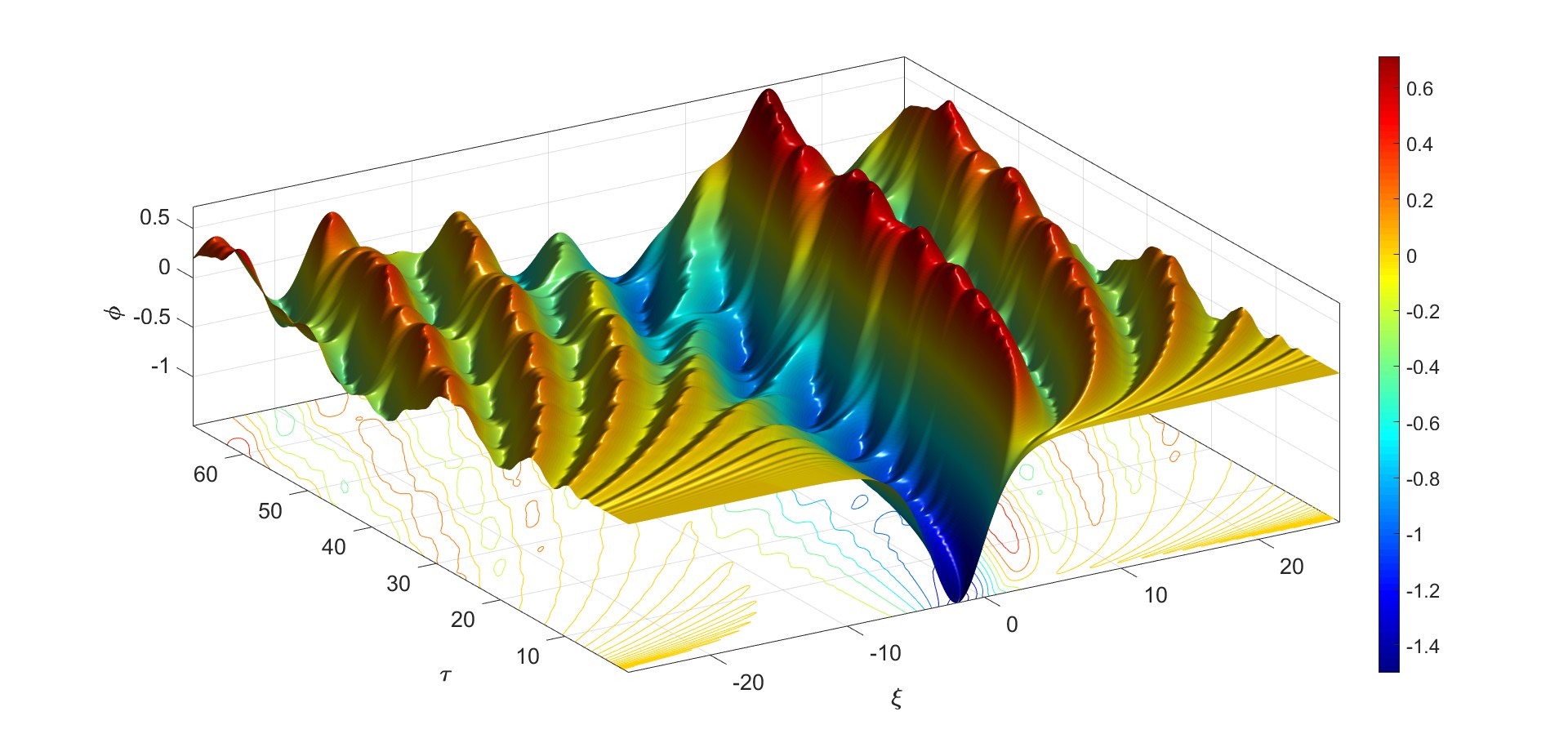} \label{C_mkdv_2a}} \hspace{0.2 in}
\subfigure[]{\includegraphics[width=0.42\linewidth]{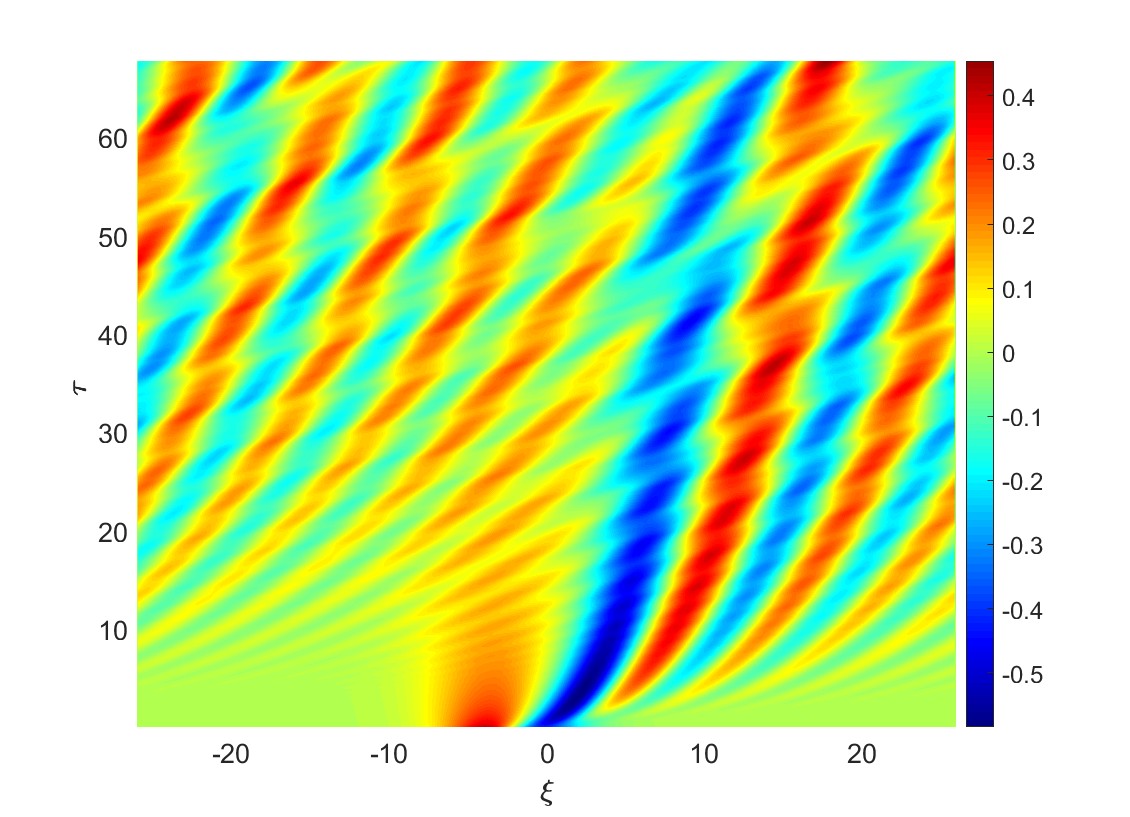} \label{C_mkdv_2b}} 
\caption{(a) Potential profile and (b) Field contours of numerical solution of mKdV equation with negative amplitude single soliton as initial profile, by considering $\mu_e=0.7,\mu_i=-0.6,\mu_ph=0.1,s= 1,p=1$.}
\end{figure}

\begin{figure}[H]
\centering
\subfigure[]{\includegraphics[width=0.42\linewidth]{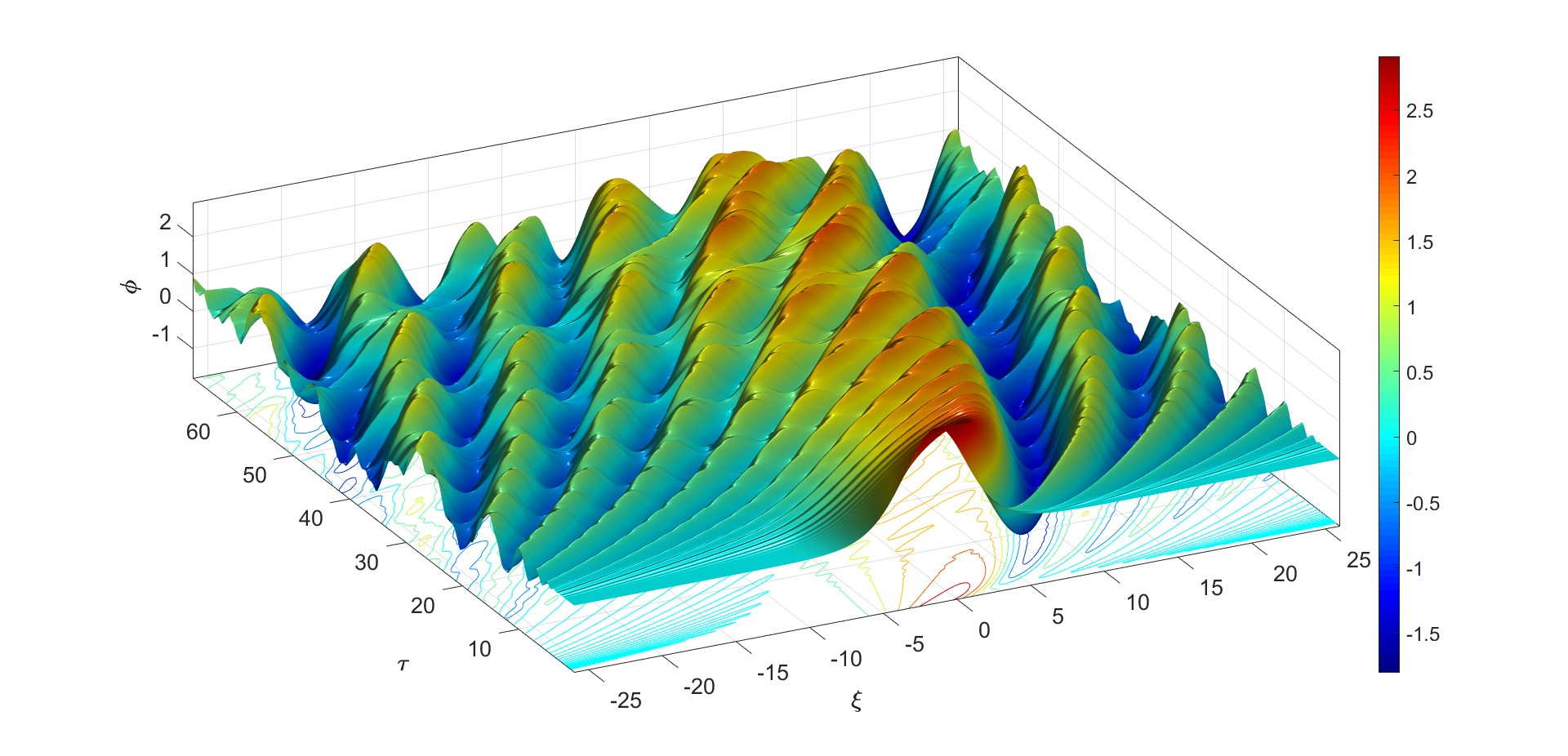} \label{C_mkdv_3a}} \hspace{0.2 in}
\subfigure[]{\includegraphics[width=0.42\linewidth]{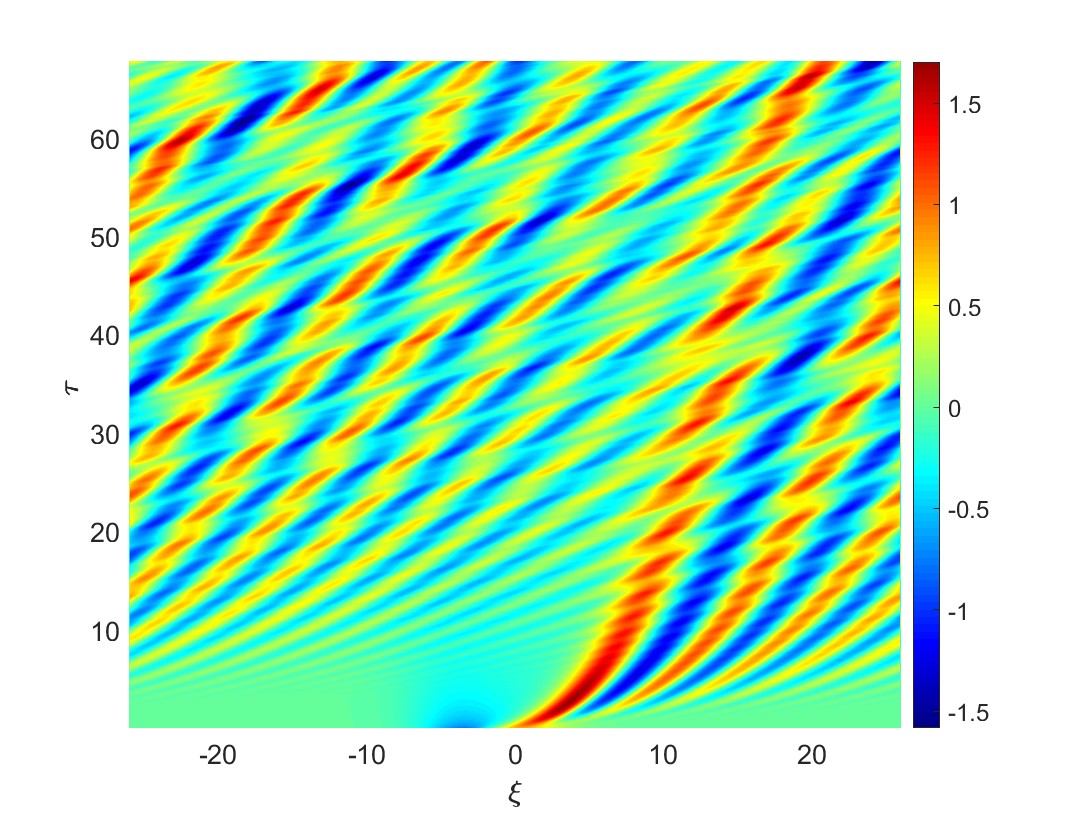} \label{C_mkdv_3b} }
\caption{(a) Potential profile and (b) Field contours of numerical solution of mKdV equation with positive amplitude double solitons as initial profile, by considering $\mu_e=0.7,\mu_i=-0.6,\mu_ph=0.1,s= 1,p=1$.}
\end{figure}
	
\begin{figure}[H]
\centering
\subfigure[]{\includegraphics[width=0.42\linewidth]{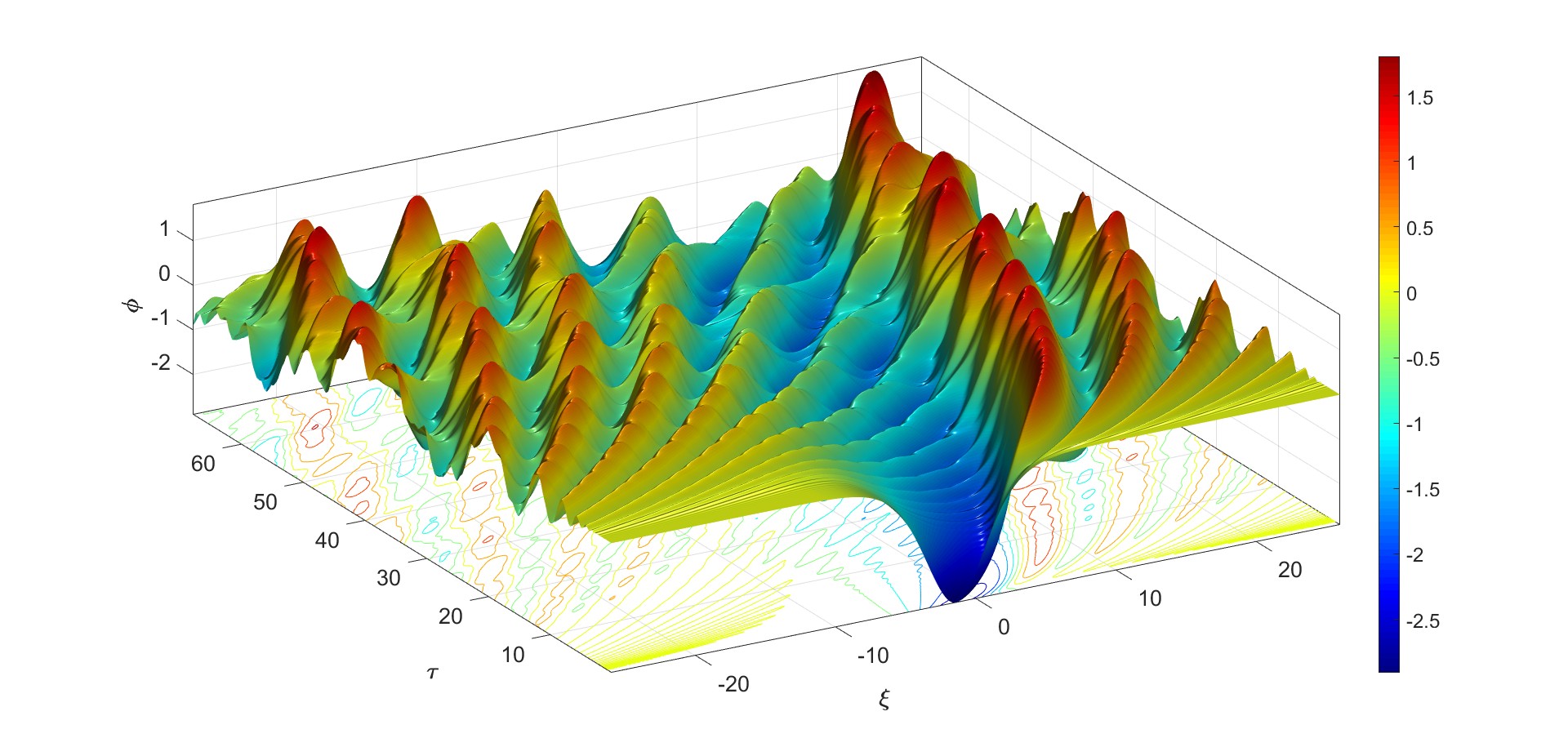} \label{C_mkdv_4a}} \hspace{0.2 in}
\subfigure[]{\includegraphics[width=0.42\linewidth]{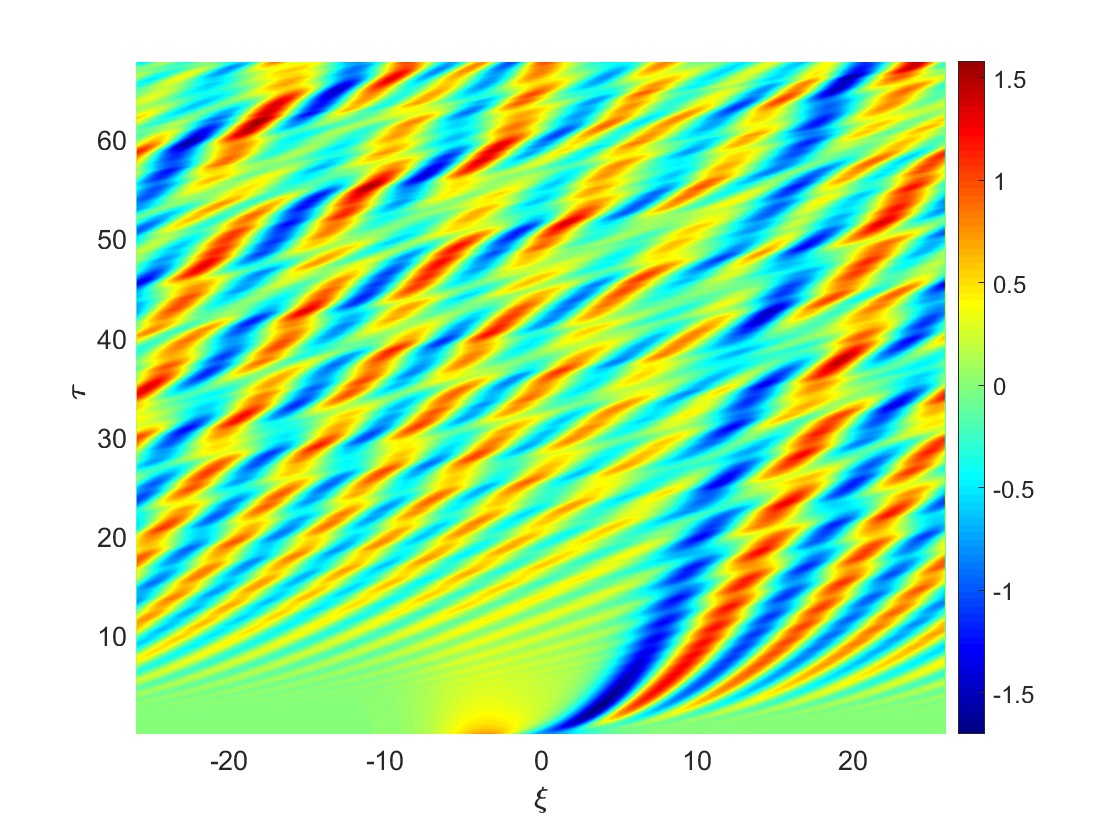} \label{C_mkdv_4b} }
\caption{(a) Potential profile and (b) Field contours of numerical solution of mKdV equation with negative amplitude double solitons as initial profile, by considering $\mu_e=0.7,\mu_i=-0.6,\mu_ph=0.1,s= 1,p=1$.}
\end{figure}
	
\begin{figure}[H]
\centering
\subfigure[]{\includegraphics[width=0.42\linewidth]{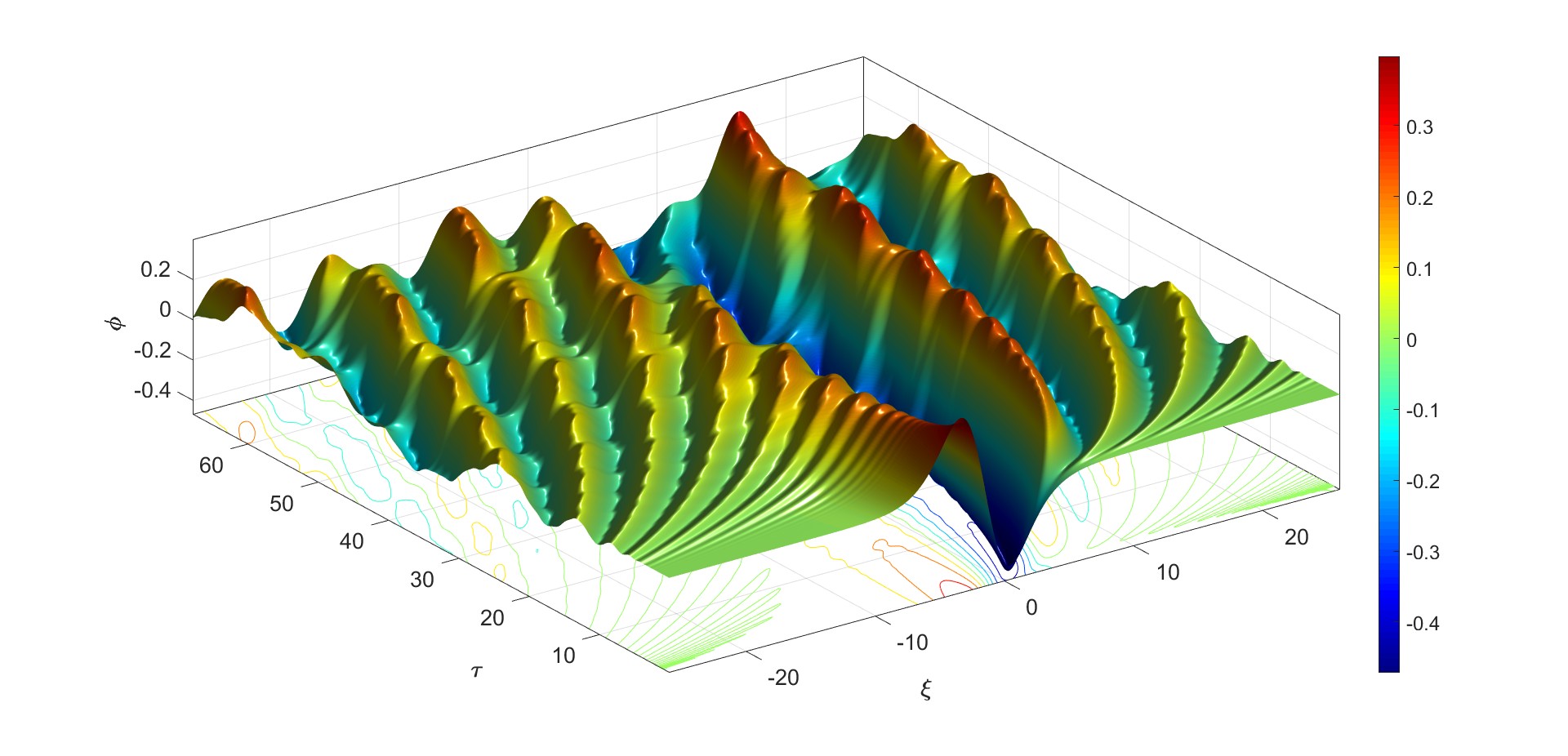} \label{C_mkdv_5a}} \hspace{0.2 in}
\subfigure[]{\includegraphics[width=0.42\linewidth]{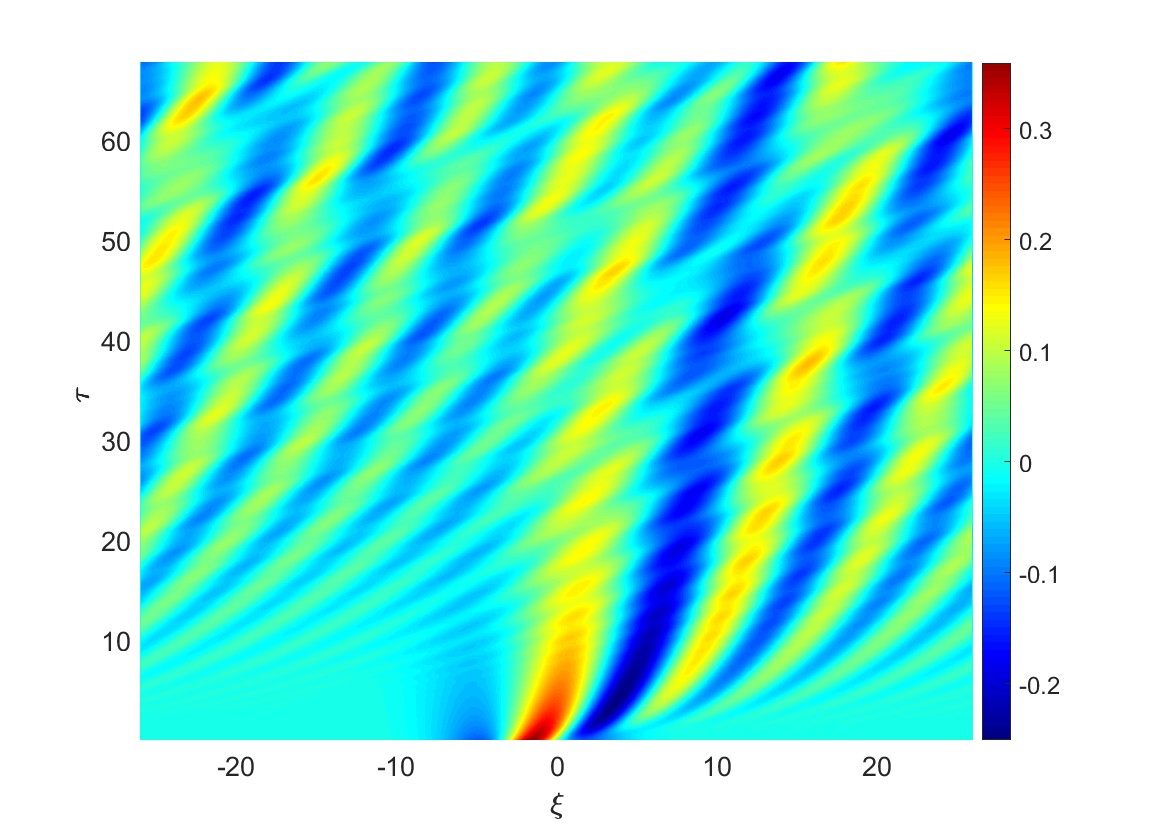} \label{C_mkdv_5b} }
\caption{(a) Potential profile and (b) Field contours of numerical solution of mKdV equation with positive and negative amplitude double solitons as initial profile, by considering $\mu_e=0.7,\mu_i=-0.6,\mu_ph=0.1,s= 1,p=1$.}
\end{figure}
	
\begin{figure}[H]
\centering
\subfigure[]{\includegraphics[width=0.42\linewidth]{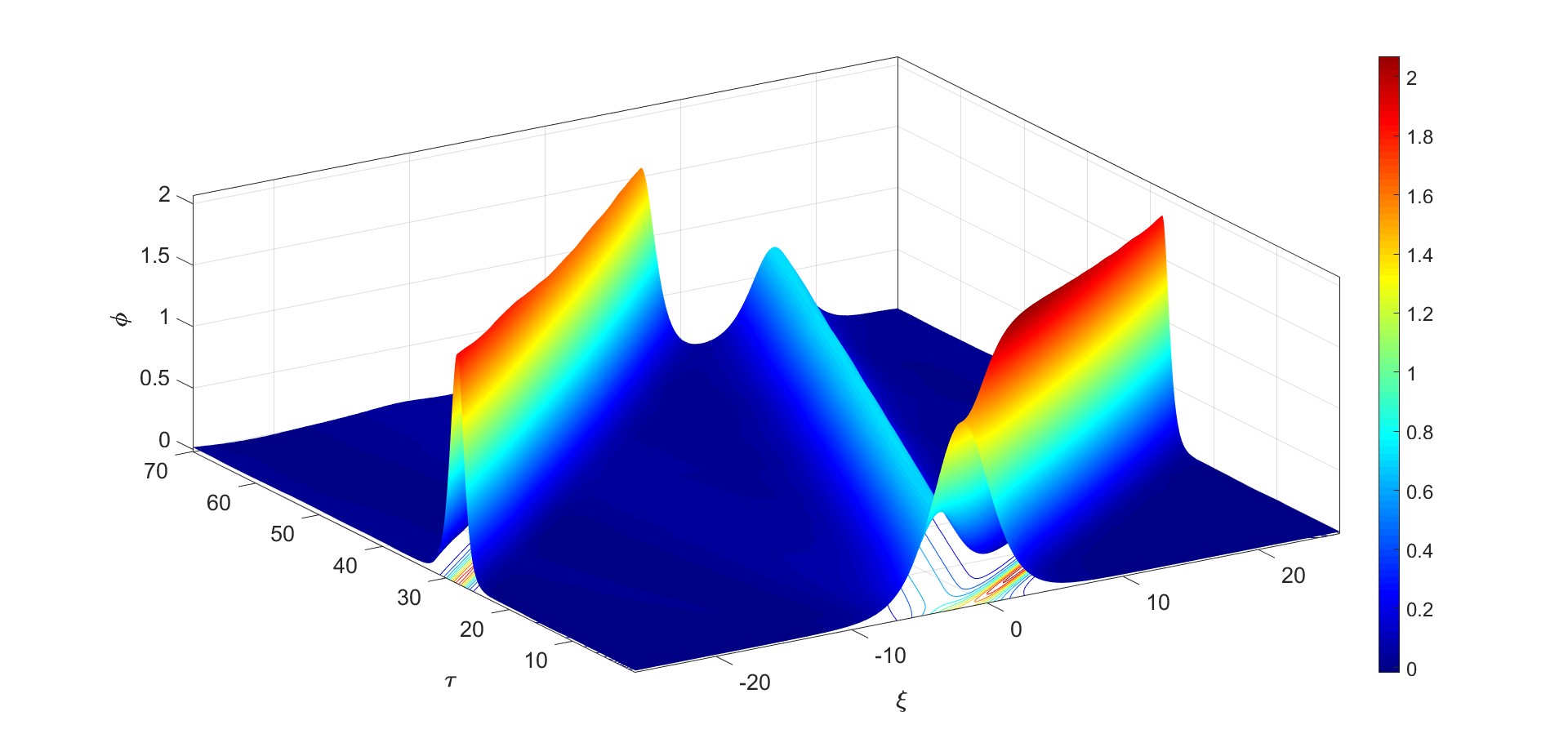} \label{C_G_1a}} \hspace{0.2 in}
\subfigure[]{\includegraphics[width=0.42\linewidth]{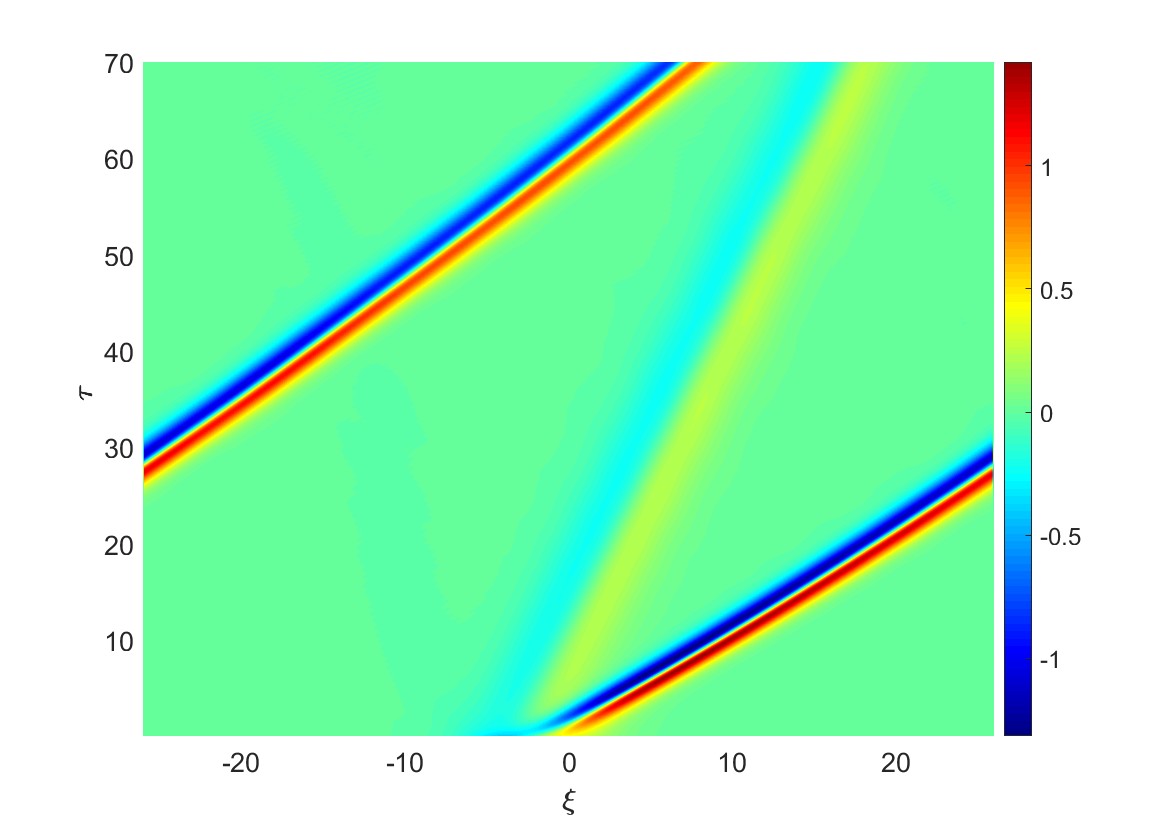} \label{C_G_1b} }
\caption{(a) Potential profile and (b) Field contours of numerical solution of Gardner equation with positive amplitude single soliton as initial profile, by considering $\mu_e=0.5,\mu_i=-0.6,\mu_ph=0.1,s= 1,p=1$.}
\end{figure}
	
\begin{figure}[H]
\centering
\subfigure[]{\includegraphics[width=0.42\linewidth]{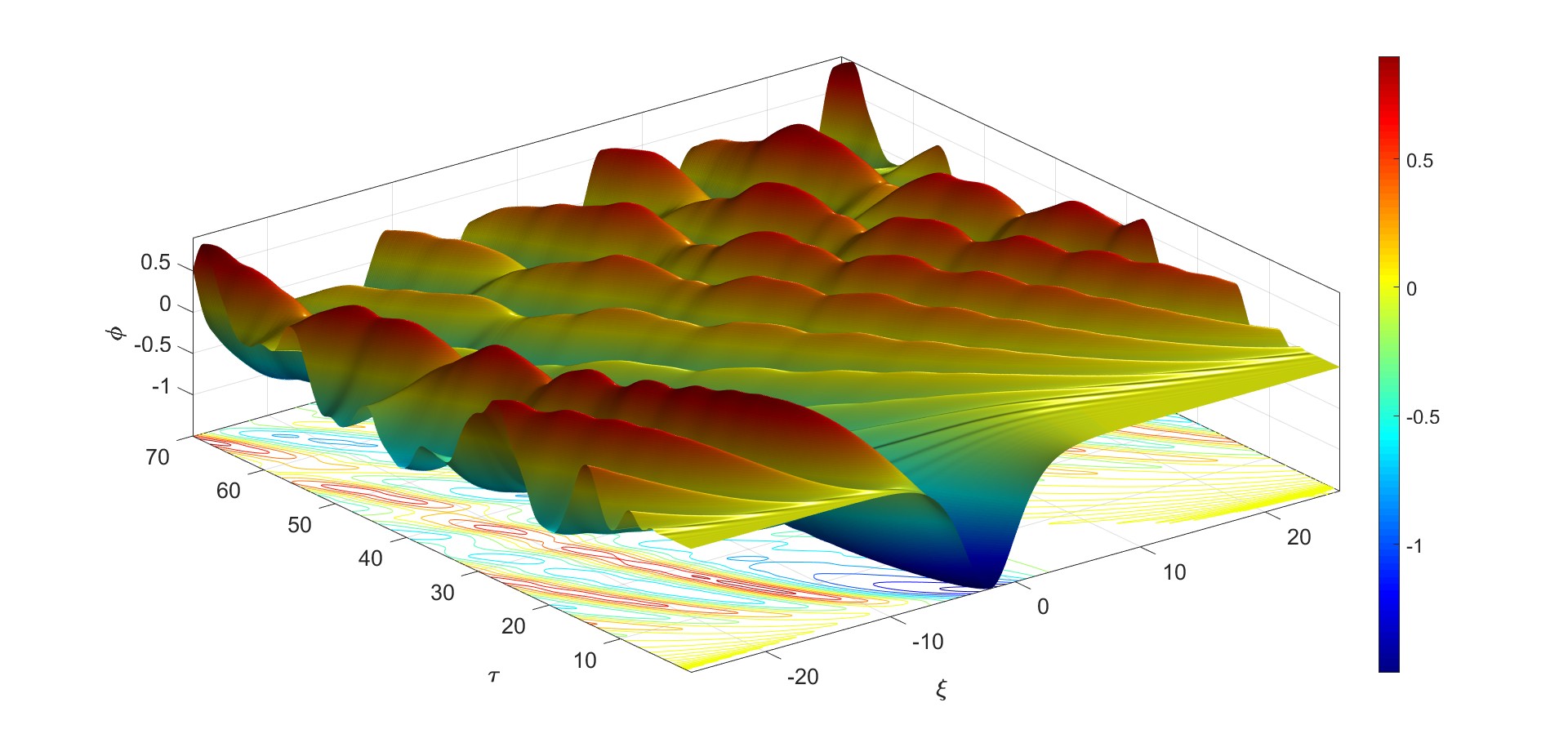} \label{C_G_2a}} \hspace{0.2 in}
\subfigure[]{\includegraphics[width=0.42\linewidth]{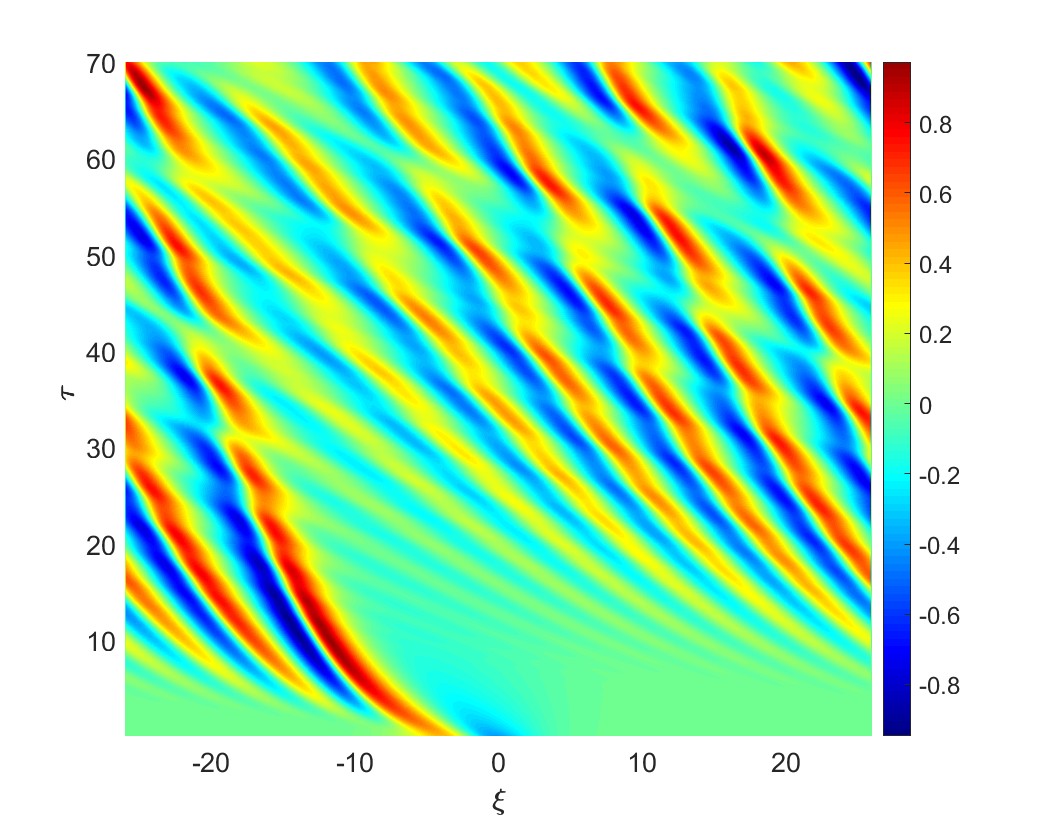} \label{C_G_2b} }
\caption{(a) Potential profile and (b) Field contours of numerical solution of Gardner equation with negative amplitude single soliton as initial profile, by considering $\mu_e=0.5,\mu_i=-0.6,\mu_ph=0.1,s= 1,p=1$.}
\end{figure}
	
\begin{figure}[H]
\centering
\subfigure[]{\includegraphics[width=0.42\linewidth]{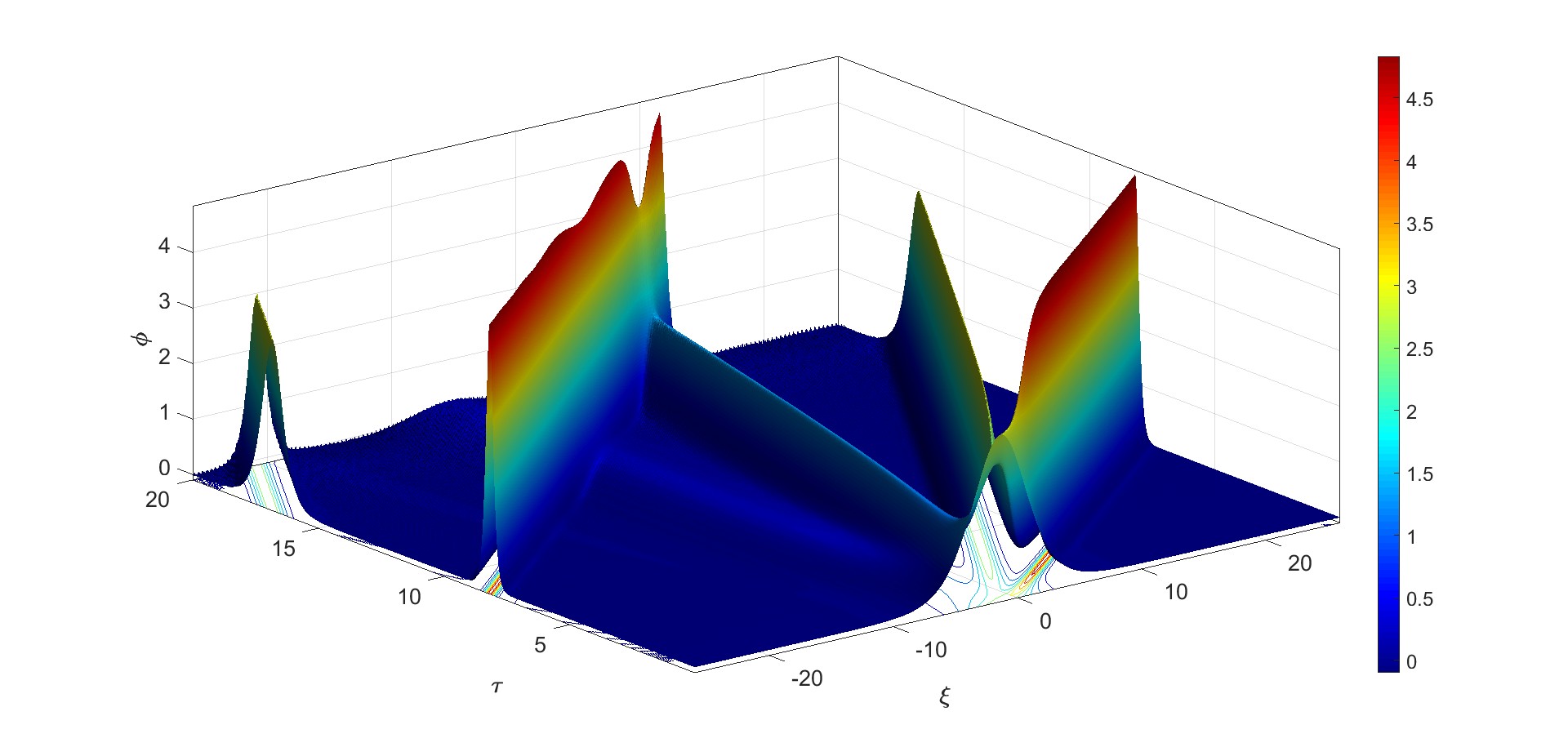} \label{C_G_3a} }\hspace{0.2 in}
\subfigure[]{\includegraphics[width=0.42\linewidth]{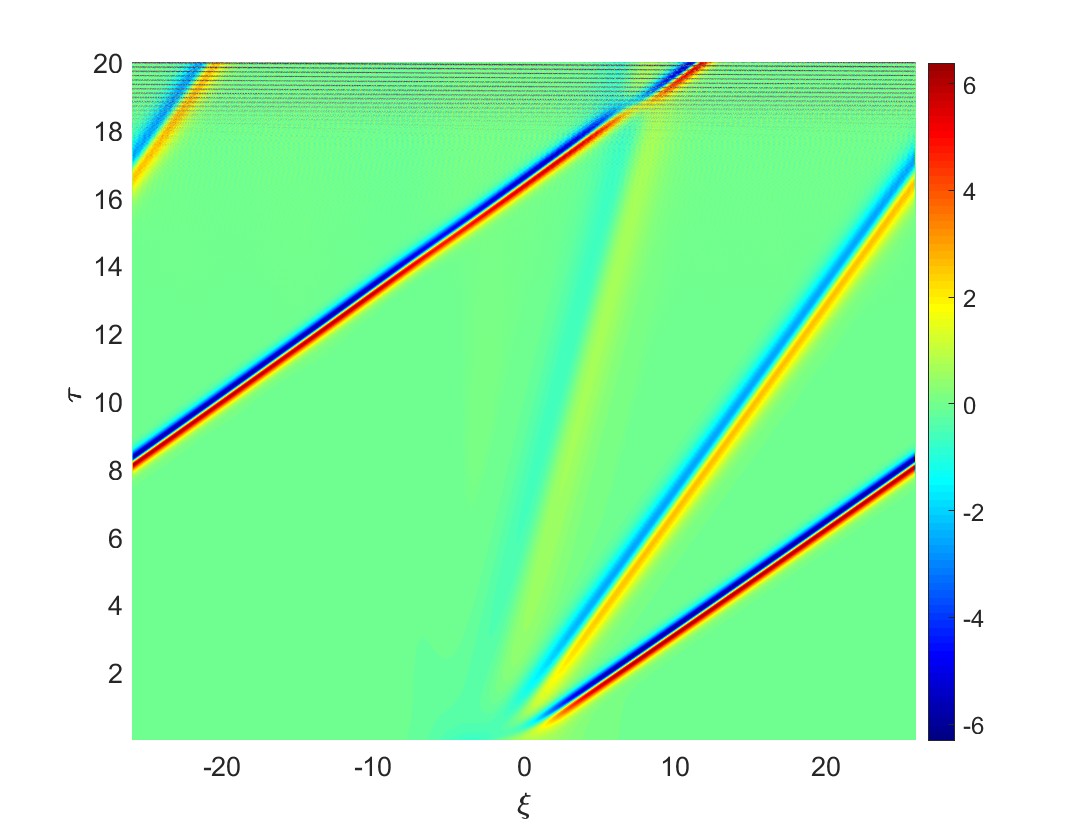} \label{C_G_3b} }
\caption{(a) Potential profile and (b) Field contours of numerical solution of Gardner equation with positive amplitude double solitons as initial profile, by considering $\mu_e=0.5,\mu_i=-0.6,\mu_ph=0.1,s= 1,p=1$.}
\end{figure}
	
\begin{figure}[H]
\centering
\subfigure[]{\includegraphics[width=0.42\linewidth]{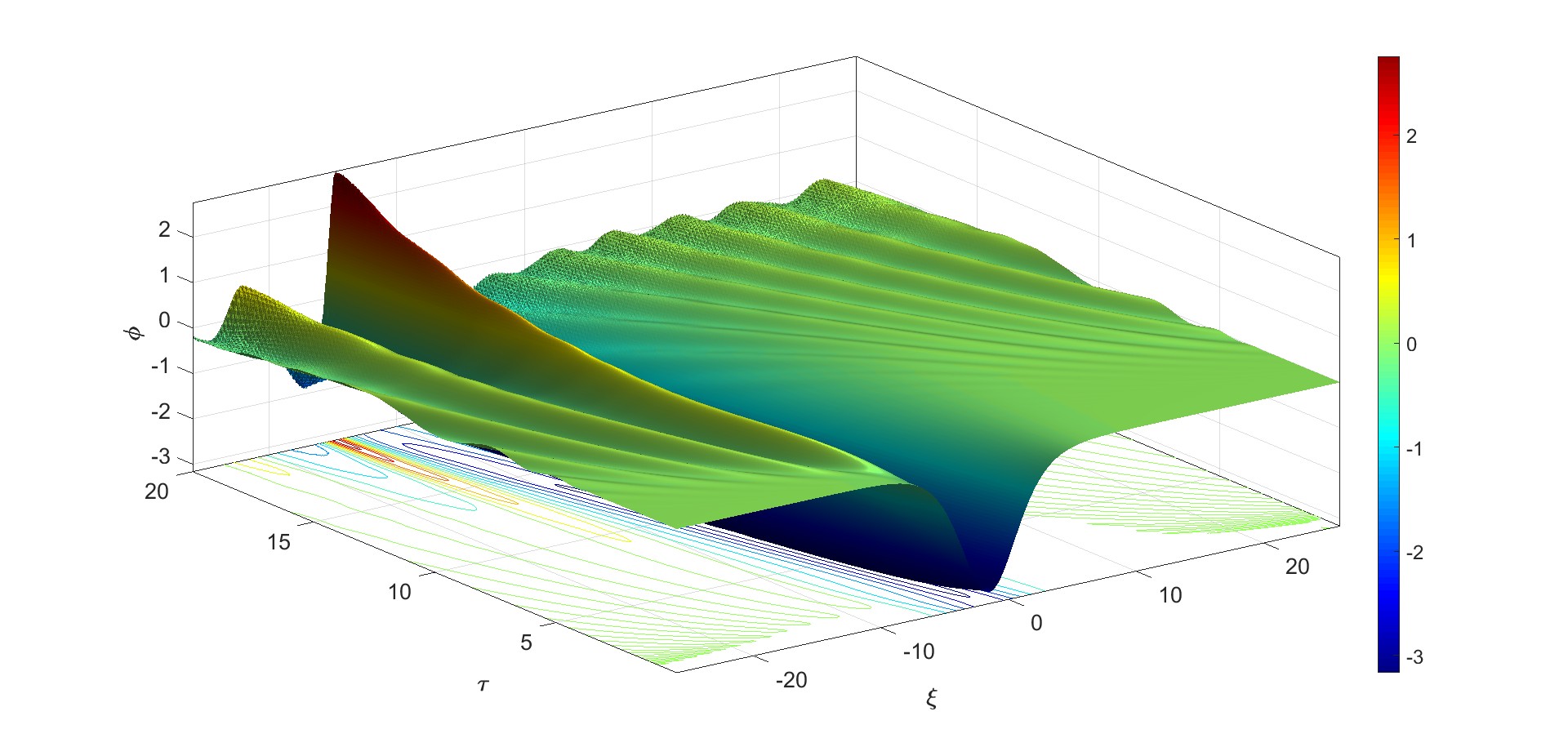} \label{C_G_4a} }\hspace{0.2 in}
\subfigure[]{\includegraphics[width=0.42\linewidth]{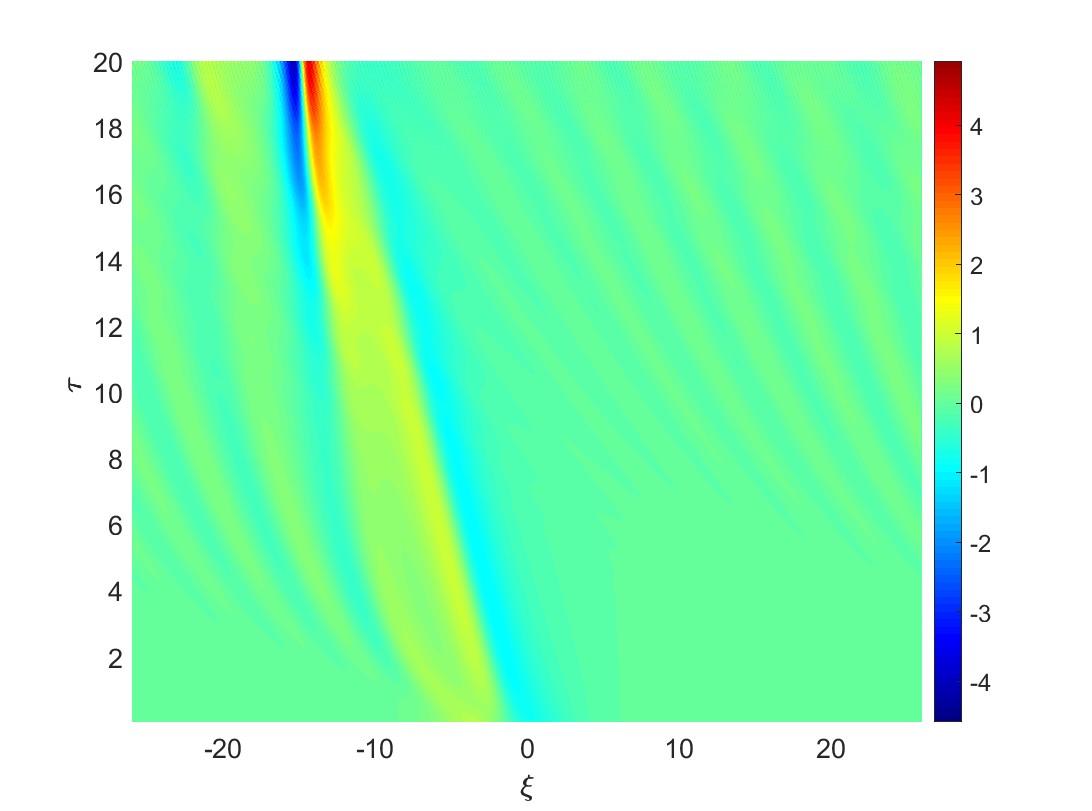} \label{C_G_4b} }
\caption{(a) Potential profile and (b) Field contours of numerical solution of Gardner equation with negative amplitude double solitons as initial profile, by considering $\mu_e=0.5,\mu_i=-0.6,\mu_ph=0.1,s= 1,p=1$.}
\end{figure}
	
\begin{figure}[H]
\centering
\subfigure[]{\includegraphics[width=0.42\linewidth]{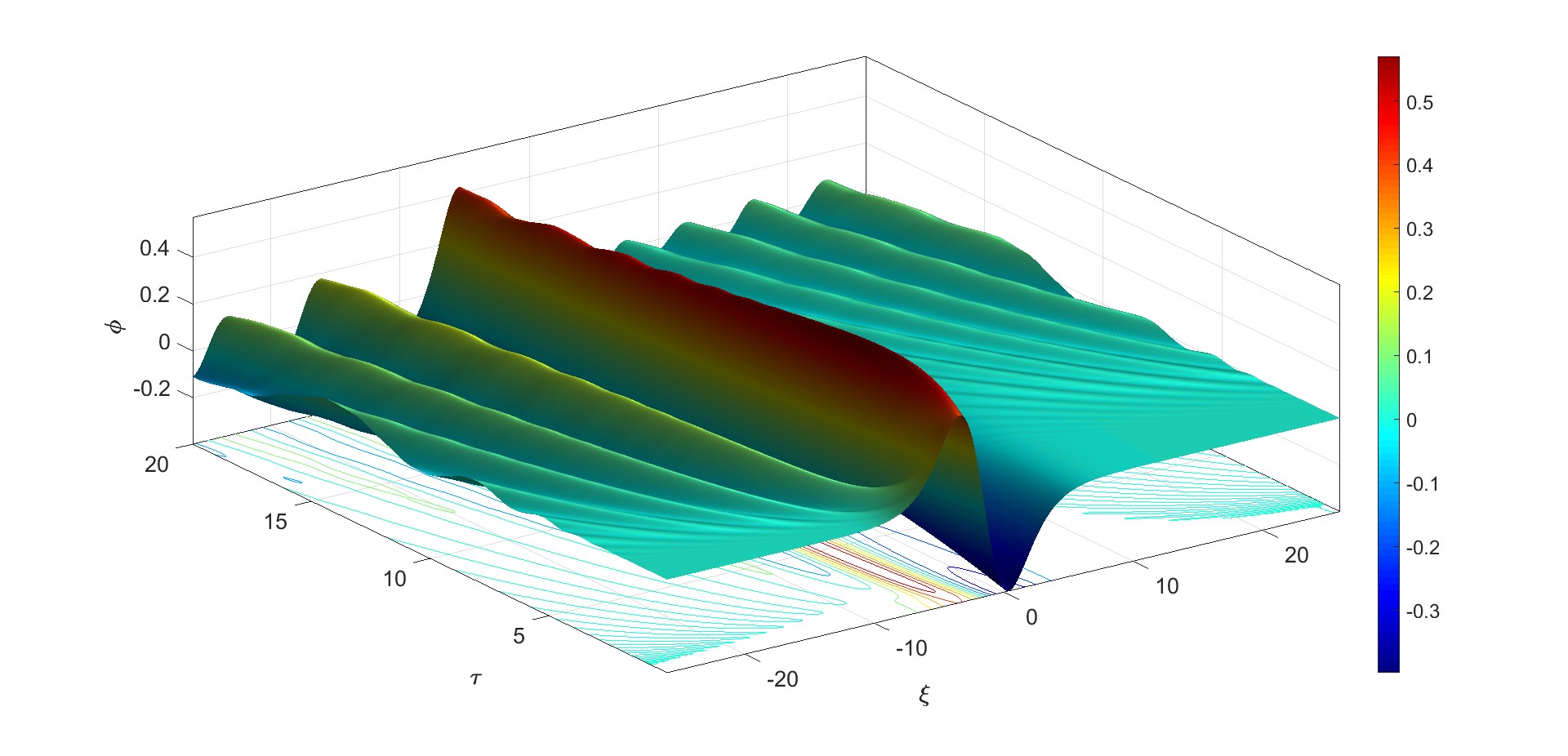} \label{C_G_5a} }\hspace{0.2 in}
\subfigure[]{\includegraphics[width=0.42\linewidth]{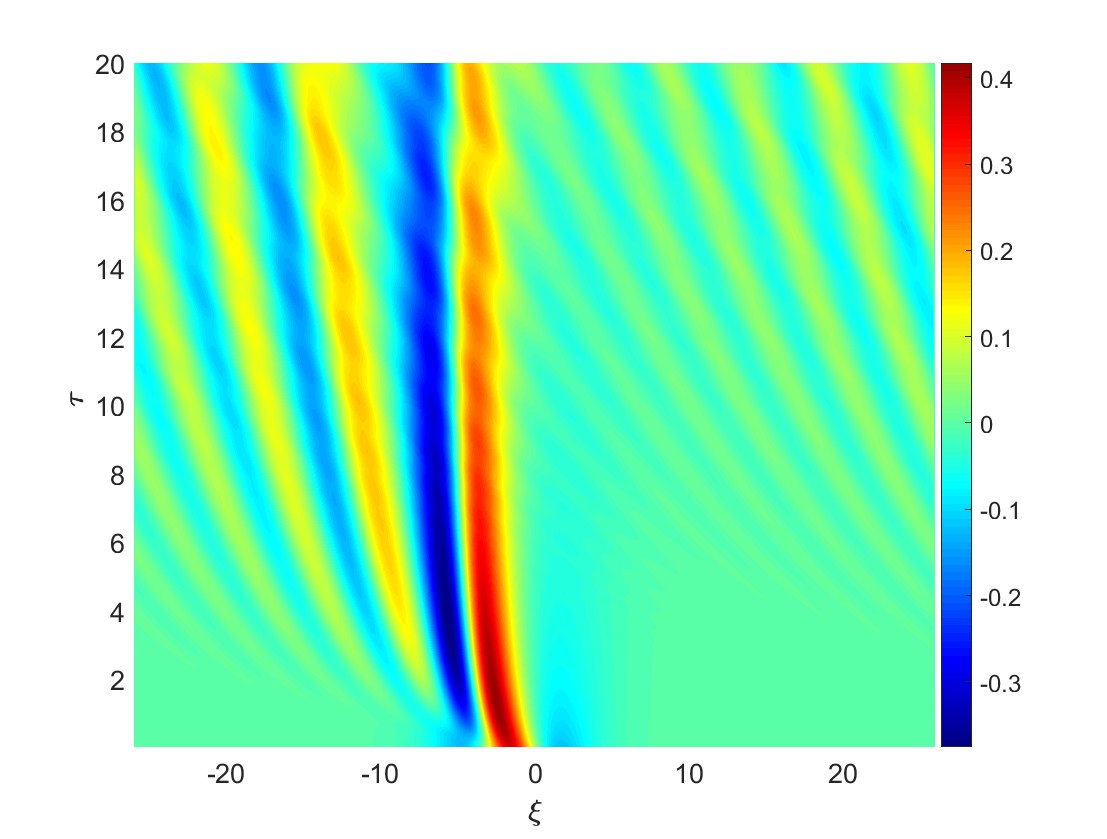} \label{C_G_5b} }
\caption{(a) Potential profile and (b) Field contours of numerical solution of Gardner equation with positive and negative amplitude double solitons as initial profile, by considering $\mu_e=0.5,\mu_i=-0.6,\mu_ph=0.1,s= 1,p=1$.}
\end{figure}
	
\section{Results and Summary}\label{results}
\subsection{Analytical Solutions:} \label{results_analy}
In the previous section, we derived the KdV and mKdV equations, and then for a critical case, we obtained the Gardner Equation (GE) \eqref{GE2}. From the analytical approach, GE can have either basic 1-soliton solution, basic 2-soliton solutions, or even a breather solution. These solutions are important from two points of view. Firstly the solutions are analytically obtained. Hence they are proof of mathematical soundness. Secondly, they can be related to similar types of findings in astrophysical and laboratory graphs and interpreted accordingly. Here we have used the parameter values $\kappa_p=100.41; \kappa_e=1.8; \sigma_1=0.8242; \sigma_2=0.5; \mu_e=0.5; \mu_i=-0.6$. For this set of parameters, we get the value of the coefficient of the quadratic nonlinear term of GE i.e. $A_1=1.87039 \times 10^{-8},$ which tends to $0$, so we consider a certain neighborhood around this parameter set, and get a cubic nonlinear term. For any case, the value of the dispersive coefficient is $B_3(=B_2=B_1)=0.396821.$  From the equilibrium equation we obtain $\mu_{ph}=0.1$. In the neighborhood where $A_1 \to 0$, we obtain the critical value of $\mu_{ph}$ i.e. $\mu_c=-3.03917$, so $\mu_{ph} > \mu_c$. Hence $s$ will be $+1$, which appear in $A_3(=c_1 s B_2)$. We have seen that the soliton solutions as well as breathers, do not depend on quadratic coefficient $A_1$, it depends on the expression $c_1=\frac{\partial A_1}{\partial \mu_{ph}}$ and its value is $c_1=2.5902$.   
	
\par{From} the analytical sections we see that the dispersive term for KdV, as well as mKdV equations, are the same (i.e. $B_1=B_2=B_3$). These are two types of nonlinearity viz. first-order nonlinearity and second-order nonlinearity. While the former (1st-order nonlinear term) i.e. $A_3$ is a composite factor given by $A_3=c_1 s B_2=c_1 s B_1$. Here $B_1(\text{or}~ B_2)$ is the dispersive term of the KdV (or mKdV) equation. Thus the first-order nonlinearity ($A_3$) depends on a variety of parameters $\sigma_1, \sigma_2, \mu_{ph}, \mu_c, \kappa_e , \kappa_p $.
\par Figures \ref{1a_1solitoncase1_2d1} \& \ref{1b_1solitoncase1_3d1} and \ref{2a_1solitoncase1_2d2} \& \ref{2b_1solitoncase1_3d2} shows oppositely propagating 1-soliton solution for the GE. It shows that the compressive soliton simply propagates without any change in amplitude or width. For phase factor ($p$) being positive, it propagates as a rarefactive soliton. Comparing figures \ref{2a_1solitoncase1_2d2} \& \ref{2b_1solitoncase1_3d2} and  \ref{3a_1solitoncase2_2d} \& \ref{3b_1solitoncase2_3d} for change in value of phase factor ($p$), variable changes appear. The rarefactive soliton remains unchanged. Next, in Figures 4 to 7, 2-soliton solutions are investigated. There are two phase factors ($p_1~ \&~ p_2$) respectively. These figures ($4-7$) provide four combinations of phase factors. When both $p_1~\&~p_2$ are positive (Figures \ref{4a_2soliton_2d1} \& \ref{4b_2soliton_3d1}) the 2-soliton approach towards the right merge into a single one, and the resultant is an overlap of the same and crosses over according to wave speeds. Figures \ref{5a_2soliton_2d2} \& \ref{5b_2soliton_3d2} and \ref{6a_2soliton_2d3} \& \ref{6b_2soliton_3d3} shows counter propagating solitons in 2-soliton mode. While the shorter one moves left $(p_1<0)$ the larger one moves right $(p_2>0)$. But in this case, the phase difference remains constant over time since the phase factors are time-independent. Such a situation can be physically observed in multi-mode optical fiber with a higher standard of preventive mechanisms towards crosstalk. An interesting occurrence happens when both $p_1~\&~p_2$ are negative (figures\ref{7a_2soliton_2d4} \& \ref{7b_2soliton_3d4}). If the analogy were drawn from figure \ref{4a_2soliton_2d1} \& \ref{4b_2soliton_3d1}, it would have been like crossing ones, however, it does not seem to cross over. These remain almost constant phase differences over time. From the \ref{7b_2soliton_3d4} figure, it is seen, the changes are very minor.
\par{We} now look for the breather-type solution when periodic modulations are seen in both space and time. A breather solution is named after its periodic contraction and expansion. There are several breather in the literature like Peregrine breathers (\cite{peregrine1},\cite{peregrine2},\cite{peregrine3} ), Ma breathers (\cite{ma1},\cite{ma2},\cite{ma3}), Akhmediev breathers (\cite{akhmediev1},\cite{akhmediev2},\cite{akhmediev3},\cite{akhmediev4}) etc. to name a few. Here a new type of breather has been presented. Starting with complex phase factors $(p_s)$ i.e. $p_1=m+in~ \& ~p_2=m-in$ where $p_1~\&~p_2$ are complex conjugate suggesting an initial correlation for positive and negative values of $m~\&~n$ and $p_1~\&~p_2$ have four combinations. \\ Figure \ref{8a_breather_2d2}\& \ref{8b_breather_3d2} shows when both $m~\&~n$ are positive. Here the movement of the breather pulse is right in the space axis suggesting it is a moving breather. The nature remains almost constant over time. \\ In figure \ref{9a_breather_2d2}\& \ref{9b_breather_3d2} we see that for both $m<0~,~n<0$, the potential edges of the breather are not equally positioned i.e. the two boundaries of the potential wall are not in the same height. This is due to the inertia of the particles and their response to change due to a propagating wave and the time taken to rest back in their original position. It seems there is some type of drag that causes a restitution of energy from the field to the particles. Figure  \ref{10a_breather_2d2}\& \ref{10b_breather_3d2} shows similar nature only the other wall height increases when compared to figure  \ref{9a_breather_2d2}\& \ref{9b_breather_3d2}. It is a clear indication that the breathers have a duration in time for which they exist and after which they break down to a shock-like structure where there is an infinitely huge potential drop. In such a region, particles are accelerated due to this huge potential difference. In figure  \ref{11a_breather_2d2}\& \ref{11b_breather_3d2} we show that the breather structure moves towards left. The rest of nature is the same. An identity inference can be drawn from figure  \ref{12a_breather_2d1}\& \ref{12b_breather_3d1}. From the values of $m~\&~n$ which are very close and oppositely signed, the breather solution has more of a shock-like features. Such an accelerated motion of the breather soliton is clearly seen from \ref{12b_breather_3d1}. To sum up this part of the result we see that starting from the GE three types of solution are obtained. These solutions help understand the gradual evolution of breather structures under the framework of GE in such an e-p-i plasmas.
\subsection{Numerical Solutions:} \label{results_num}
Thus far we have dealt with analytical solutions and investigated their nature under different conditions of positive and negative phase factors. Since analytical solutions for KdV and mKdV equations are available in the literature; for a change we try with numerical solutions for these. To obtain the numerical solutions we resort to the "Fourier Operated fourth-order Runge Kutta via Exponential Time Difference scheme" (FORKET) code and employ our Matlab platform to obtain the results. The result part is organized in the following way; in the first section, we present the figures corresponding to analytical solutions containing 2-D (field contours) and 3-D (potential profiles) plots of 1-soliton, 2-soliton and breather solutions for various parameters. The 3-D figures have time as the third axis whereas the 2-D ones have it as assuming parameter. From the figures, we see that negative values of time ($\tau$) have been shown. This has nothing to do with physicality since we choose the epoch reference somewhere in between. This part of the numerical scheme was done by two of the authors (SC \& CD).  Through this scheme, we have plotted the potential structures ($\phi$) and the fields ($E$) for KdV, mKdV, and GE. The plots of KdV and mKdV will help us to understand the intermediate process. The numerical scheme is very much more fine-tuned than other RK4 schemes. Therefore, with parametric variations, the growth of KdV, mKdV, and Gardner solitons can be better realized through this tool. 
\par The parameters for numerical plots are given in the caption. First, we study the evolution of the KdV solitary profile and the electric field as well. Figure (\ref{C_kdv_1a}) shows the solitonic evolution from the KdV equation. The KdV soliton propagates towards the right. However, there are secondary modulations that propagate in the opposite direction. Figure (\ref{C_kdv_1b}) shows the corresponding electric field in the $\xi-\tau$ plane. Subsequent figures will also show similar features. A thorough study of this feature suggests two possible mechanisms.\\ (i) When positive ions form the ion acoustic wave (IAW), the density pile up forces the heavier ions to stay back which start small amplitude oscillations about their mean position thus creating these secondary modulations.\\ (ii) Alternatively, as ions pile up to form IAW, positrons in the plasma are repelled back thus forming low amplitude waves which start propagating within the plasma in the reverse direction to the direction of propagation of IAW solitons. \\ When the initial amplitude is taken negative (Figure \ref{C_kdv_2a}) i.e. ions are depleted from the soliton resulting in an excess of super-thermal electrons. These super-thermal electrons create crocodilian back-like secondary humps. These were not visible in previous cases. However the field nature (figure \ref{C_kdv_2b}) is similar except for the sharp intense field that was existing in figure \ref{C_kdv_1a}.  In figures \ref{C_kdv_3a} \& \ref{C_kdv_3b}, two solitonic solutions have been started with. The initial nature is a hyperbolic secant squared type. The two solitons have different phase speeds and it is clear from the curved nature of the path of the soliton that the phase speed is not constant over time, rather a nonlinearity exists itself in the phase. This curvature however negligible is proof that the monochromatic concept of solitary wave is not valid completely. Now if there are two solitons, one compressive and the other rarefactive (figure \ref{C_kdv_4a} \& \ref{C_kdv_4b}), then a train of secondary modulations are seen, these are deep as well a shallow pockets of particle trapping. The corresponding field diagram shows the deep potential walls marked with darker shades of blue. Under certain critical conditions where the nonlinear term of ordinary KdV evolution is zero, then we take a step forward to obtain the mKdV equation. We start with a single soliton case and see that both spatially and temporally modulated solitons are formed and often the single soliton gets deformed and forms a series of secondary solitons (figure \ref{C_mkdv_1a}). The corresponding field (figure \ref{C_mkdv_1b}) shows the interaction of primary and secondary solitons. If the initial structure is that of a rarefactive soliton, the potential deep gradually becomes less shallow, and secondary solitons are compressive. Here also the phase does depend on time in a nonlinear fashion (figure \ref{C_mkdv_2a}). The corresponding fields figure \ref{C_mkdv_2b} are presented through color code where local sources and sinks appear either as a continuum or sporadically referring to small-scale oscillations. The figures (\ref{C_mkdv_3a} \& \ref{C_mkdv_3b}) and (\ref{C_mkdv_4a} \& \ref{C_mkdv_4b}) we start with 2-soliton solutions of different phase speeds. In the first case, both are positive compressive solitons whereas in the latter case, both are negative rarefactive solitons. These two pairs of figures show the intermittent crests and troughs from which we infer that field lines emerge and converge throughout the bulk of the plasma. The contour of equipotential surfaces at the base of potential profiles clearly second the conclusion we drew from.. the field directions. The picturesque structures (figure \ref{C_mkdv_5a}), where rarefactive and other compressive solitons are taken to be the 2-soliton solutions of the mKdV equation. Low amplitude solitary profiles grow progressively showing alterations both in space ($\xi$) and time ($\tau$). The fields (figure \ref{C_mkdv_5b}) are however not as strong as the previous two cases. This is visible from the color codes (light green and light blue being predominant). 
\par Next the numerical solutions for the Gardner equation are given. The spatio-temporal variations in potential profiles and field contours show the breather mode to move away from its origin along the phase axis for a single soliton (figures \ref{C_G_1a} \& \ref{C_G_1b}). If given sufficient time, the breather would appear again at a different position, where the phase factor matches. If the polarity is reversed striated cloud-like features appear and it is not any breather mode, rather a train of successive modulations appears (figures \ref{C_G_2a} \& \ref{C_G_2b}). If two soliton solutions both compressive (figures \ref{C_G_3a} \& \ref{C_G_3b}) or rarefactive (figures \ref{C_G_4a} \& \ref{C_G_4b}) are studied, positive solitons evolution are different from those of negative soliton. The possibility of breather modes start appearing. This is rather interesting since IAWs are formed due to ions, and the positive ones' evolution is understood. However, with initial negative solitons the secondary modulations are due to mobile positrons, which are super-thermally Kappa distributed. This is also reported in the analytic investigation. Lastly, we study the case for numerical solution, where one compressive and another rarefactive, regular undulations appear and contours are elongated channels rather than pockets of crests and troughs in the time axis. Thus through this numerical technique, we were able to study how the IAW transfers from KdV solitons to Gardner breathers or solitons in detail.
	
\section{Conclusion} \label{concl}
The work presented so far contains analytic and numerical studies of KdV, mKdV, and GE. The derivations are done in detail. Analytic solutions were used along with numerical solutions. The findings of the analytic solutions of GE for 1-soliton, 2-solitons, or breathers are presented in detail. Numerically, all three types of equations were studied. The importance of this work is twofold. From the mathematical modeling to analytic solutions this work presents a plethora for future work and can be considered as a theoretician's playground. From the applicability part to real physics, the usefulness covers astrophysics to Laser plasma physics, and even semiconductor plasma where electron-holes and ions have similar fluid equations. The high energetic particles often called super-thermal in the literature follow the Kappa distribution. We have surveyed a lot of literature and found that a non-Maxwellian distribution is present in many space and laboratory plasma environments. Depending upon the composition and configurations the physical problem in a way to suit an astrophysical environment. Going for the techniques we have employed a sophisticated numerical scheme (ETDRK4) which does add merit to our work. We have observed that in a certain parameter region both the quadratic and cubic nonlinearities exist (in GE) and balancing with dispersion produces interesting soliton structures i.e. breather structures for a four-component plasma system consisting of immobile positive ions, mobile cold positrons, and Kappa distributed hot positrons and hot electrons. No investigation has been done related to this topic. Many researchers have studied such models but they haven't investigated the breather structures in such plasma systems. We first establish the existence of breathers for GE in a four-component plasma system consisting of immobile positive ions. mobile cold positrons and Kappa distributed hot positrons and hot electrons.

\section*{Acknowledgment}
Ms.Snehalata Nasipuri is thankful to the Council of Scientific and Industrial Research (CSIR), India, for providing financial support under the Senior Research Fellowship (SRF) program (File No: 09/202(0120)/2021-EMR-I). \\\\
\section*{Author Contributions}
S.N. calculated the analytical part, plotted the results, and wrote the text. P.C. supervised the whole work. U.N.G.  modelled  and analysed the results, also wrote some parts of this article. All authors have discussed to the final manuscript.\\\\
\textbf{Competing interests and fundings}\\
No funding was received to assist with the preparation of this manuscript. Also, the author has no financial or proprietary interests in any material discussed in this article.\\\\
\textbf{Data availability statement}\\
The author confirms that there is no associated data available for the above research work. Data sharing does not
apply to this article as no new data were created or analyzed in this study.\\\\
{\large{\textbf{Declarations}}}	
\textbf{Conflict of Interest}\\
The author has no conflict of interest

\end{document}